\documentclass[11pt,fullpage]{article}

\usepackage{times,amsmath,epsfig}
\usepackage{graphicx}
\usepackage{amsmath,amssymb,amstext}
\usepackage{subfig}
\usepackage{paralist}
\usepackage{caption}
\usepackage{url}
\usepackage{algorithm,algorithmic}
\usepackage{multirow}
\usepackage{placeins}
\usepackage{fullpage}

\newcommand{\eat}[1]{}

\newcommand{\E}[1]{E\left[#1\right]}
\newcommand{\Var}[1]{\text{Var}\left(#1\right)}

\newcommand{\EstVar}[1]{\text{Est}_{#1}}

\newtheorem{lema}{Lemma}

\begin{document}
\date{February 2013}

\title{PF-OLA: A High-Performance Framework for Parallel Online Aggregation}

\author{
%
%
Chengjie Qin \hspace*{2cm} Florin Rusu\\
       \small{University of California, Merced}\\
       \small{5200 N Lake Road}\\
       \small{Merced, CA 95343}\\
       \small\texttt{\{cqin3,frusu\}@ucmerced.edu}
}

\maketitle

\begin{abstract}

Online aggregation provides estimates to the final result of a computation during the actual processing. The user can stop the computation as soon as the estimate is accurate enough, typically early in the execution. This allows for the interactive data exploration of the largest datasets.

In this paper we introduce the first framework for parallel online aggregation in which the estimation virtually does not incur any overhead on top of the actual execution. We define a generic interface to express any estimation model that abstracts completely the execution details. We design a novel estimator specifically targeted at parallel online aggregation. When executed by the framework over a massive $8\text{TB}$ TPC-H instance, the estimator provides accurate confidence bounds early in the execution even when the cardinality of the final result is seven orders of magnitude smaller than the dataset size and without incurring overhead.

\end{abstract}

\section{Introduction}\label{sec:intro}

Interactive data exploration is a prerequisite in model design. It requires the analyst to execute a series of exploratory queries in order to find patterns or relationships in the data. In the Big Data context, it is likely that the entire process is time-consuming even for the fastest parallel database systems given the size of the data and the sequential nature of exploration---the next query to be asked is always dependent on the previous. Online aggregation~\cite{ola,ola-sigrec} aims at reducing the duration of the exploration process by allowing the analyst to rule out the non-informative queries early in the execution. To make this possible, an estimate to the final result of the query with progressively narrower confidence bounds is continuously returned to the analyst. When the confidence bounds become tight enough -- typically early in the processing -- the analyst can decide to stop the execution and focus on the next query.

Although introduced in the late nineties, it is hard to find any commercial system that supports online aggregation even today~\cite{AQP-book}. In our opinion, there are multiple reasons that hindered adoption given that multiple academic prototypes~\cite{control,demo:dbo} proved the feasibility of the approach. The first argument underlines the negative effect online aggregation has on normal query execution, with increases of at least $30\%$ being common~\cite{turbo:dbo,ripple-join}. This is regarded as unacceptable in the commercial world who focus instead on improving the performance of traditional database systems. The second argument addresses the lack of a unified approach to express estimation models. The authors went multiple times through the process of designing a new estimation model in previous projects~\cite{turbo:dbo}. Each time, major changes to the overall system architecture and a significant amount of implementation were required. These obstructed the ultimate goal of designing a better estimation model. The final argument against online aggregation adoption we mention is the requirement to re-implement the data processing system from ground up in order to support estimation.

PF-OLA (\textbf{P}arallel \textbf{F}ramework for \textbf{O}n\textbf{L}ine \textbf{A}ggregation) overpasses these limitations and brings online aggregation closer to broader adoption, especially for Big Data interactive exploration. PF-OLA is a shared nothing parallel system executing complex analytical queries over terabytes of data very efficiently. Estimates and corresponding confidence bounds for the query result are computed during the entire query execution process without incurring any noticeable overhead. Thus, a user executing a long-running query starts getting estimates with provable guarantees almost instantly after query processing starts. As the query progresses, the width of the confidence bounds shrinks progressively, converging to the true result when the query is complete. As far as we know, PF-OLA is the first system that incurs virtually no overhead on top of the query execution time corresponding to the non-interactive execution. This is due to the extensive use of parallelism at all levels of the system -- including storage, inside a single node, and across all the nodes -- and to a judicious overlapping of query execution and estimation. At the same time, neither the estimator accuracy nor the convergence rate are negatively impacted, but rather the convergence receives a significant boost from the parallel discovery of result tuples. This results in fast and accurate estimations even for highly selective queries with very low result cardinality -- the needle in the haystack problem -- or in the case of skewed data.

A second aspect that differentiates PF-OLA from other online aggregation systems is the unified approach to express estimation models. In PF-OLA, the estimation is closely intertwined with the actual computation, with both clearly separated from query execution. Any computation is expressed as a User-Defined Aggregate (UDA)~\cite{uda}. The user is responsible for implementing a standard interface while the framework takes care of all the issues related to parallel execution and data flow. Adding online estimation to the computation is a matter of extending the UDA state and interface with constructs corresponding to the estimator. In order to apply a different estimation model to a computation, a corresponding UDA has to be implemented. No changes have to be made to the implementation of the computation or to the implementation of the framework. This provides tremendous flexibility in designing a variety of estimation models.

To verify the expressiveness of the framework and test the performance, we design an asynchronous sampling estimator specifically targeted at parallel online aggregation. The estimator is defined over multiple data partitions which can be independently sampled in parallel. This allows for accurate estimates to be computed even when there is considerable variation between the sampling rate across partitions. We analyze the properties of the estimator and compare it with two other sampling-based estimators proposed for parallel online aggregation---a synchronous estimator~\cite{distributed-ola} and a stratified sampling estimator~\cite{scalable-hash-ripple}. All these estimators are expressed using the extended UDA interface and executed without any changes to the framework, thus proving the generality of our approach.

The complete system re-implementation is avoided in PF-OLA through the use of the generic UDA mechanism to express both computation as well as estimation. The framework defines only the execution strategy without imposing any limitation on the actual computation. As long as the user-provided computation and estimation model can be expressed using the generic mechanism, there is no need to change the PF-OLA implementation. And, as shown in~\cite{GLADE-OSR,bismarck}, the complexity of the tasks that can be expressed with the UDA mechanism ranges from simple and group-by aggregations to clustering and convex optimization problems applied in machine learning.

Our main contributions can be summarized as follows:
\begin{compactitem}
\item We design the first framework for parallel online aggregation that incurs virtually no overhead on top of the actual execution. Estimates and corresponding confidence bounds are continuously computed based on samples extracted from data during the entire processing.
\item We define a generic interface to express estimation models by extending the well-known UDA mechanism. The user is required to represent the model using a pre-defined set of methods while the framework handles all the execution details in a parallel environment.
\item We implement the online aggregation framework in a highly-parallel processing system for the execution of arbitrary jobs. The result is an extremely efficient prototype for Big Data analytics with support for online aggregation.
\item We propose a novel asynchronous sampling estimator for parallel online aggregation that we implement and execute inside the framework. We provide statistical analysis, verify the correctness, and show the superior performance when compared with two other estimators previously proposed in the literature.
\item We run an extensive set of experiments to benchmark the estimator. When executed by the framework over a massive $8\text{TB}$ TPC-H instance~\cite{tpch}, the estimator provides accurate confidence bounds early in the execution even when the cardinality of the result is seven orders of magnitude smaller than the dataset size or when data are skewed without incurring any noteworthy overhead on top of the normal execution. Moreover, the estimator exhibits high resilience in the wake of processing node delays and failures.
\end{compactitem}

\paragraph{Roadmap.}
In the remainder of the paper, we first introduce a set of preliminary concepts in Section~\ref{sec:prelim}. Parallel online aggregation is formalized in Section~\ref{sec:par-online-agg} which contains a detailed presentation of our proposed estimator and a thorough comparison with existent estimators. The design of the framework and the implementation details are discussed in Section~\ref{sec:pf-ola}. Example estimators and their implementation in PF-OLA are presented in Section~\ref{sec:est-examples}, while Section~\ref{sec:empirical} contains the empirical evaluation of the framework and of the proposed estimator. Related work is discussed in Section~\ref{sec:rel-work}. We conclude by summarizing the main findings of this paper and providing future directions in Section~\ref{sec:conclusions}.

\section{Preliminaries}\label{sec:prelim}

We consider aggregate computation in a parallel cluster environment consisting of multiple processing nodes. Each processing node has a multi-core processor consisting of one or more CPUs, thus introducing an additional level of parallelism. Data are partitioned into fixed size chunks that are stored across the processing nodes. Parallel aggregation is supported by processing multiple chunks at the same time both across nodes as well as across the cores inside a node.

We focus on the computation of general \texttt{SELECT-PROJECT-JOIN} (SPJ) queries having the following SQL form~\cite{turbo:dbo}:
\begin{equation}\label{eq:prob-def}
\begin{split}
&\texttt{SELECT SUM(f($\texttt{t}_{\texttt{1}}$ $\bullet$ $\texttt{t}_{\texttt{2}}$))}\\
&\texttt{FROM $\texttt{TABLE}_{\texttt{1}}$ AS $\texttt{t}_{\texttt{1}}$, $\texttt{TABLE}_{\texttt{2}}$ AS $\texttt{t}_{\texttt{2}}$}\\
&\texttt{WHERE P($\texttt{t}_{\texttt{1}}$ $\bullet$ $\texttt{t}_{\texttt{2}}$)}
\end{split}
\end{equation}
where $\bullet$ is the concatenation operator, \texttt{f} is an arbitrary \textit{associative decomposable aggregate function}~\cite{GLADE-OSR} over the tuple created by concatenating $\texttt{t}_{\texttt{1}}$ and $\texttt{t}_{\texttt{2}}$, and \texttt{P} is some boolean predicate that can embed selection and join predicates. The class of associative decomposable aggregate functions, i.e., functions that are associative and commutative, is fairly extensive and includes the majority of standard SQL aggregate functions. Associative decomposable aggregates allow for the maximum degree of parallelism in their evaluation since the computation is independent of the order in which data inside a chunk are processed as well as of the order of the chunks, while partial aggregates computed over different chunks can be combined together straightforwardly. While the paper does not explicitly discuss aggregate functions other than \texttt{SUM}, functions such as \texttt{COUNT}, \texttt{AVERAGE}, \texttt{STD DEV}, and \texttt{VARIANCE} can all be handled easily---they are all associative decomposable. For example, \texttt{COUNT} is a special case of \texttt{SUM} where \texttt{f}($\cdot$) = 1 for any tuple, while \texttt{AVERAGE} can be computed as the ratio of \texttt{SUM} and \texttt{COUNT}.

\texttt{GROUP BY} queries can also be handled using the methods in this paper by simply treating each group as a separate query and running all queries simultaneously; then all of the estimates are presented to the user. For each group, a version of \texttt{P} is used that accepts only tuples from that particular group.

\subsection{Parallel Aggregation}\label{sec:par-agg}

Aggregate evaluation takes two forms in parallel databases. They differ in how the partial aggregates computed for each chunk are combined together. In the centralized approach, all the partial aggregates are sent to a common node -- the coordinator -- that is further aggregating them to produce the final result. As an intermediate step, local aggregates can be first combined together and only then sent to the coordinator. In the parallel approach, the nodes are first organized into an aggregation tree. Each node is responsible for aggregating its local data and the data of its children. The process is executed level by level starting from the leaves, with the final result computed at the root of the tree. The benefit of the parallel approach is that it also parallelizes the aggregation of the partial results across all the nodes rather than burdening a single node (with data and computation). The drawback is that in the case of a node failure it is likely that more data are lost. Notice that these techniques are equally applicable inside a processing node, at the level of a multi-core processor.

PF-OLA supports both centralized and parallel aggregation at the level of the entire cluster as well as inside each node. The strategy to be applied is determined dynamically for each query. Moreover, when online aggregation is executed simultaneously with the normal query execution, the aggregation strategy is chosen individually for each of the tasks. Thus, for example, it is possible to have the query executed with parallel aggregation, while the estimation is centralized.

\subsection{Online Aggregation}\label{sec:online-agg}

The idea in online aggregation is to compute only an estimate of the aggregate result based on a sample of the data~\cite{ola}. In order to provide any useful information though, the estimate is required to be accurate and statistically significant. Different from one-time estimation~\cite{AQP-book,TamingTerabytes} that might produce very inaccurate estimates for arbitrary queries, online aggregation is an iterative process in which a series of estimators with improving accuracy are generated. This is accomplished by including more data in estimation, i.e., increasing the sample size, from one iteration to another. The end-user can decide to run a subsequent iteration based on the accuracy of the estimator. Although the time to execute the entire process is expected to be much shorter than computing the aggregate over the entire dataset, this is not guaranteed, especially when the number of iterations is large. Other issues with \textit{iterative online aggregation}~\cite{distributed-ola,earl} regard the choice of the sample size at each iteration and reusing the work done from one iteration to the following.

An alternative that avoids these problems altogether is to completely \textit{overlap query processing with estimation}~\cite{demo:dbo,online-mapreduce}. As more data are processed towards computing the final aggregate, the accuracy of the estimator improves accordingly. For this to be true though, data are required to be processed in a statistically meaningful order, i.e., random order, to allow for the definition and analysis of the estimator. This is typically realized by randomizing data during the loading process. The drawback of the overlapped approach is that the same query is essentially executed twice---once towards the final aggregate and once for computing the estimator. As a result, the total execution time in the overlapped case is expected to be higher when compared to the time it takes to execute each task separately.

PF-OLA is designed as an online aggregation system which overlaps query execution with estimation. The motivation for this choice is the ever increasing number of cores available on modern CPUs. Since I/O is the bottleneck in database processing, the additional computation power is not utilized unless concurrent tasks are found and executed. Given that the estimation process requires access to the same data as normal processing, estimation is a natural candidate for overlapped execution. While it is straightforward to execute these two processes concurrently, the challenge is how to realize this such that the normal query execution time is not increased at all---this should always be possible as long as there are non-utilized cores on the processing node. We show how PF-OLA achieves this goal by carefully scheduling access to shared data across the two tasks.

\section{Parallel Online Aggregation}\label{sec:par-online-agg}

An online aggregation system provides estimates and confidence bounds during the entire query execution process. As more data are processed, the accuracy of the estimator increases while the confidence bounds shrink progressively, converging to the actual query result when the entire data have been processed. There are multiple aspects that have to be considered in the design of a parallel online aggregation system. First, a mechanism that allows for the computation of partial aggregates has to be devised. Second, a parallel sampling strategy to extract samples from data over which partial aggregates are computed has to be designed. Each sampling strategy leads to the definition of an estimator for the query result, estimator that has to be analyzed in order to derive confidence bounds. We discuss in details each of these aspects for the overlapped online aggregation approach in this section. Then, in Section~\ref{sec:pf-ola}, we show how everything is implemented in the PF-OLA framework.

\subsection{Partial Aggregation}\label{ssec:par-online-partial-agg}

The first requirement in any online aggregation system is a mechanism to compute partial aggregates over some portion of the data. Partial aggregates are typically a superset of the query result since they have to contain additional data required for estimation. The partial aggregation mechanism can take two forms. We can fix the subset of the data used in partial aggregation and execute a normal query. Or we can interfere with aggregate computation over the entire dataset to extract partial results before the computation is completed. The first alternative corresponds to iterative online aggregation, while the second to overlapped execution.

Partial aggregation in a parallel setting raises some interesting questions. For iterative online aggregation, the size and location of the data subset used to compute the partial aggregate have to be determined. It is common practice to take the same amount of data from each node in order to achieve load balancing. Or to have each node process a subset proportional to its data as a fraction from the entire dataset. Notice though that it is not necessary to take data from all the nodes. In the extreme case, the subset considered for partial aggregation can be taken from a single node. Once the data subset at each node is determined, parallel aggregation proceeds normally, using either the centralized or parallel strategy. In the case of overlapped execution, a second process that simply aggregates the current results at each node has to be triggered whenever a partial aggregate is computed. The aggregation strategy can be the same or different from the strategy used for computing the final result. Centralized aggregation might be more suitable though due to the reduced interference. The amount of data each node contributes to the result is determined only by the processing speed of the node. Since the work done for partial aggregation is also part of computing the final aggregate, it is important to reuse the result so that the overall execution time is not increased unnecessarily.

\subsection{Parallel Sampling}\label{ssec:par-online-sampling}

In order to provide any information on the final result, partial aggregates have to be statistically significant. It has to be possible to define and analyze estimators for the final result using partial aggregates. Online aggregation imposes an additional requirement. The accuracy of the estimator has to improve when more data are used in the computation of partial aggregates. In the extreme case of using the entire dataset to compute the partial aggregate, the estimator collapses on the final result. The net effect of these requirements is that the data subset on which the partial aggregate is computed cannot be arbitrarily chosen. Since sampling satisfies these requirements, the standard approach in online aggregation is to choose the subset used for partial aggregation as a random sample from the data.

\paragraph{Centralized sampling.}
Thus, an important decision that has to be taken when designing an online aggregation system is how to generate random samples. According to the literature~\cite{olken-phd}, there are two methods to generate samples from the data in a centralized setting. The first method is based on using an index that provides the random order in which to access the data. While it does not require any pre-processing, this method is highly inefficient due to the large number of random accesses to the disk. The second method is based on the idea of storing data in random order on disk such that a sequential scan returns random samples at any position. Although this method requires considerable pre-processing at loading time to permute data randomly, it is the preferred randomization method in online aggregation systems~\cite{continuous-sampling} since the cost is paid only once and it can be amortized over the execution of multiple queries---the indexing method incurs additional cost for each query. As a result, PF-OLA implements a parallel version of the random shuffling method.

\paragraph{Sampling synopses.}
It is important to make the distinction between the runtime sampling methods used in online aggregation and estimation based on static samples taken offline~\cite{AQP-book,TamingTerabytes}, i.e., sampling synopses. In the later case, a sample of fixed size is taken only once and all subsequent queries are answered using the sample. This is typically faster than executing sampling at runtime, during query processing. The problem is that there are queries that cannot be answered from the sample accurately enough, for example, highly selective queries. The only solution in this case is to extract a larger sample entirely from scratch which is prohibitively expensive. The sampling methods for online aggregation described previously  avoid this problem altogether due to their incremental design that degenerates in a sample consisting of the entire dataset in the worst case.

\paragraph{Sample size.}
Determining the correct sample size to allow for accurate estimations is an utterly important problem in the case of sampling synopses and iterative online aggregation. If the sample size is not large enough, the entire sampling process has to be repeated, with unacceptable performance consequences. While there are methods that guide the selection of the sample size for a given accuracy in the case of a single query, they require estimating the variance of the query estimator---an even more complicated problem. In the case of overlapped online aggregation, choosing the sample size is not a problem at all since the entire dataset is processed in order to compute the correct result. The only condition that has to be satisfied is that the data seen up to any point during processing represent a sample from the entire dataset. As more data are processed towards computing the query result, the sample size increases automatically. Both runtime sampling methods discussed previously satisfy this property.

\begin{figure*}
\centering
\subfloat[]{\includegraphics[width=0.5\textwidth]{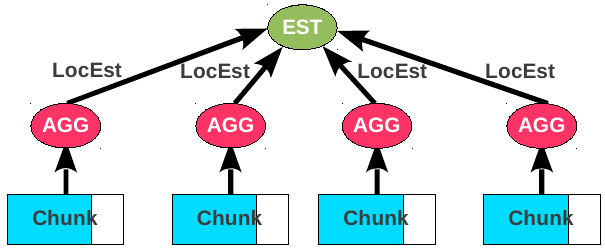}\label{fig:paragg-centr}}
\subfloat[]{\includegraphics[width=0.5\textwidth]{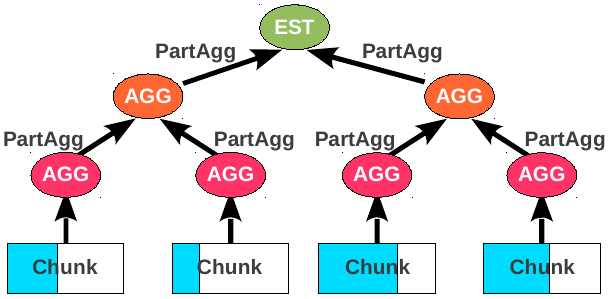}\label{fig:paragg-tree}}
\caption{(a) Centralized aggregation with multiple estimators. (b) Aggregation tree with single estimator.}\label{fig:paragg-strategy}
\end{figure*}

\paragraph{Stratified sampling.}
There are multiple alternatives to obtain a sample from a partitioned dataset---the case in a parallel setting. The straightforward solution is to consider each partition independently and to apply centralized sampling algorithms inside the partition (Figure~\ref{fig:paragg-centr}). This type of sampling is known as \textit{stratified sampling}~\cite{sampling-techniques}. While stratified sampling generates a random sample for each partition, it is not guaranteed that when putting all the local samples together the resulting subset is a random sample from the entire data. For this to be the case, it is required that the probability of a tuple to be in the sample is the same across all the partitions. The immediate solution to this problem is to take local samples that are proportional with the partition size.

\begin{algorithm}[htbp]
\caption{\textit{RandomSplit} (Dataset $D$ on node $n$)}
\label{alg:random-split}
\algsetup{linenodelimiter=.}

\textbf{Input:} Number of nodes $N$; Random hash function $h$ \\
\textbf{Output:} Partition $P = \{D_{1}, D_{2}, \dots, D_{N}\}$ of $D$

\begin{algorithmic}[1]
\FORALL {items $d \in D$}
\STATE Let $k = h(d)$
\STATE Add $d$ to set $D_{k}$ in partition $P$
\ENDFOR
\end{algorithmic}
\end{algorithm}

\paragraph{Global randomization.}
A somehow more complicated solution is to make sure that a tuple can reside at any position in any partition---\textit{global randomization} (Figure~\ref{fig:paragg-tree}). This can be achieved by randomly shuffling the data across all the nodes---as a direct extension of the similar centralized approach. The global randomization process consists of two stages, each executed in parallel at every node. In the first stage (Algorithm~\ref{alg:random-split}), each node partitions the local data into sets corresponding to all the other nodes in the environment. The assignment of an item to a partition is based on a random hash function which requires as argument a random value independent of the item in order to randomize the data. The index of the item in the dataset (round-robin partitioning) might be a good choice as long as the set assignment does not follow a predictable pattern. An even better choice is a random value generated for the item.

\begin{algorithm}[htbp]
\caption{\textit{RandomPermutation} (Data fragments from all $N$ nodes at node $n$)}
\label{alg:random-permute}
\algsetup{linenodelimiter=.}

\textbf{Input:} Set $P^{n} = \bigcup_{i=1}^{N}{D^{n}_{i}}$ of data fragments from all $N$ nodes; Random number generator \textit{rng} \\
\textbf{Output:} Random permutation of $P^{n}$

\begin{algorithmic}[1]
\FORALL {items $p \in P^{n}$}
\STATE Let $r_{p} = $\textit{rng}$(p)$ be a random number for item $p$
\ENDFOR
\STATE Sort $P^{n}$ in the increasing order of $r_{p}$
\end{algorithmic}

\end{algorithm}

In the second stage of the randomization process (Algorithm~\ref{alg:random-permute}), each node generates a random permutation of the data received from all the other nodes---random shuffling. This is required in order to separate the items received from the same origin. The standard method for random shuffling consists in generating a random value for each item in the dataset and then sorting the items according to these random values. Notice that using the random values from the origin node is not guaranteed to produce a random permutation across all the sets.

The main benefit provided by global randomization is that it simplifies the complexity of the sampling process in a highly-parallel asynchronous environment. This in turn allows for compact estimators to be defined and analyzed---a single estimator across the entire dataset. It also supports more efficient sampling algorithms that require a reduced level of synchronization, as is the case with our estimator.Moreover, global randomization has another important characteristic for online aggregation---it allows for incremental sampling. What this essentially means is that in order to generate a sample of a larger size starting from a given sample is enough to obtain a sample of the remaining size. It is not even required that the two samples are taken from the same partition since random shuffling guarantees that a sample taken from a partition is actually a sample from the entire dataset. Equivalently, to get a sample from a partitioned dataset after random shuffling, it is not necessary to get a sample from each partition.

While random shuffling in a centralized environment is a time-consuming process executed in addition to data loading, global randomization in a parallel setting is a standard hash-based partitioning process executed as part of data loading. Due to the benefits provided for workload balancing and for join processing, hash-based partitioning is heavily used in parallel data processing even without online aggregation. Thus, we argue that global randomization for parallel online aggregation is part of the data loading process and it comes at no cost with respect to sampling.

\subsection{Estimation}\label{ssec:est:process}

While designing sampling estimators for online aggregation in a centralized environment is a well-studied problem, it is not so clear how these estimators can be extended to a highly-parallel asynchronous system with data partitioned across nodes. To our knowledge, there are two solutions to this problem proposed in the literature. In the first solution, a sample over the entire dataset is built from local samples taken independently at each partition. An estimator over the constructed sample is then defined. We name this approach \textit{single estimator}. In the single estimator approach, the fundamental question is how to generate a single random sample of the entire dataset from samples extracted at the partition level. The strategy proposed in~\cite{distributed-ola} requires synchronization between all the sampling processes executed at partition level in order to guarantee that the same fraction of the data is sampled at each partition. To implement this strategy, serialized access to a common resource is required for each item processed. This results in a factor of four increase in execution time when estimation is active (see the experimental evaluation section).

In the second solution, which we name \textit{multiple estimators} (Figure~\ref{fig:paragg-centr}), an estimator is defined for each partition. As in stratified sampling theory~\cite{sampling-techniques}, these estimators are then combined into a single estimator over the entire dataset. The solution proposed in~\cite{scalable-hash-ripple} follows this approach. The main problem with the multiple estimators strategy is that the final result computation and the estimation are separated processes with different states that require more complicated implementation.

We propose an asynchronous sampling estimator specifically targeted at parallel online aggregation that combines the advantages of the existing strategies. We define our estimator as in the single estimator solution, but without the requirement for synchronization across the partition-level sampling processes which can be executed independently (Figure~\ref{fig:paragg-tree}). This results in much better execution time. When compared to the multiple estimators approach, our estimator has a much simpler implementation since there is complete overlap between execution and estimation. In this section, we analyze the properties of the estimator and compare it with the two estimators it inherits from. Then, in Section~\ref{sec:est-examples} we provide insights into the actual implementation in PF-OLA, while in Section~\ref{sec:empirical} we present experimental results to evaluate the accuracy of the estimator and the runtime performance of the estimation.

\subsubsection{Generic Sampling Estimator}\label{sssec:est-gen}

To design estimators for the parallel aggregation problem we first introduce a generic sampling estimator for the centralized case. This is a standard estimator based on sampling without replacement~\cite{sampling-techniques} that is adequate for online aggregation since it provides progressively increasing accuracy. We define the estimator for the simplified case of aggregating over a single table and then show how it can be generalized to \texttt{GROUP BY} and general SPJ queries (Equation~\ref{eq:prob-def}) in Section~\ref{sec:est-examples}.

Consider the dataset $D$ to have a single partition sorted in random order. The number of items in $D$ (size of $D$) is $|D|$. While sequentially scanning $D$, any subset $S \subseteq D$ represents a random sample of size $|S|$ taken without replacement from $D$. We define an estimator for the aggregate as follows:
\begin{equation}\label{eq:gen-est}
	X = \frac{|D|}{|S|} \sum_{s \in S, \texttt{P}(s)} \texttt{f}(s)
\end{equation}
where $X$ has the properties given in Lemma~\ref{lema:gen-est-moments}:

\begin{lema}\label{lema:gen-est-moments}
	$X$ is an unbiased estimator for the aggregation problem, i.e., $\E{X} = \sum_{d \in D, \texttt{P}(d)} \texttt{f}(d)$, where $\E{X}$ is the expectation of $X$. The variance of $X$ is equal to:
\begin{equation}\label{eq:gen-est-var}
  \begin{split}
	& \Var{X} = \frac{|D|-|S|}{(|D|-1) |S|} \left[|D| \sum_{d \in D, \texttt{P}(d)} \texttt{f}^{2}(d) - \left(\sum_{d \in D, \texttt{P}(d)} \texttt{f}(d) \right)^{2} \right] \\
  \end{split}
\end{equation}
\end{lema}

It is important to notice the factor $|D|-|S|$ in the variance numerator which makes the variance to decrease while the size of the sample increases. When the sample is the entire dataset, the variance becomes zero, thus the estimator is equal to the exact query result. The standard approach to derive confidence bounds~\cite{sms-join,dbo,turbo:dbo} is to assume a normal distribution for estimator $X$ with the first two frequency moments given by $\E{X}$ and $\Var{X}$. The actual bounds are subsequently computed at the required confidence level from the cumulative distribution function (cdf) of the normal distribution. Since the width of the confidence bounds is proportional with the variance, a decrease in the variance makes the confidence bounds to shrink.

A closer look at the variance formula in Equation~\ref{eq:gen-est-var} reveals the dependency on the entire dataset $D$ through the two sums over all the items $d \in D$ that satisfy the selection predicate \texttt{P}. Unfortunately, when executing the query we have access only to the sampled data. Thus, we need to compute the variance from the sample. We do this by defining a variance estimator, $\EstVar{\Var{X}}$, with the following properties:

\begin{lema}\label{lema:gen-est-var-est}
	The estimator
\begin{equation}\label{eq:gen-est-var-est}
  \begin{split}
	& \EstVar{\Var{X}} = \frac{|D|(|D|-|S|)}{|S|^{2}(|S|-1)} \left[|S| \sum_{s \in S, \texttt{P}(s)} \texttt{f}^{2}(s) - \left(\sum_{s \in S, \texttt{P}(s)} \texttt{f}(s) \right)^{2} \right] \\
  \end{split}
\end{equation}
	is an unbiased estimator for the variance in Equation~\ref{eq:gen-est-var}.
\end{lema}

Having the two estimators $X$ and $\EstVar{\Var{X}}$ computed over the sample $S$, we are in the position to provide the confidence bounds required by online aggregation in a centralized environment. The next step is to extend the generic estimators to the parallel setting of the PF-OLA framework where data is partitioned across multiple processing nodes.

\subsubsection{Single Estimator Sampling}\label{sssec:est-agg-one}

In the PF-OLA framework, the dataset $D$ is partitioned across $N$ processing nodes, i.e., $D = D_{1} \cup D_{2} \cup \dots \cup D_{N}$. A sample $S_{i}$, $1\leq i \leq N$, is taken independently at each node. These samples are then put together in a sample $S = S_{1} \cup S_{2} \cup \dots \cup S_{N}$ over the entire dataset $D$. To guarantee that $S$ is indeed a sample from $D$, in the case of the synchronized estimator in~\cite{distributed-ola} it is enforced that the sample ratio $\frac{S_{i}}{D_{i}}$ is the same across all the nodes. For the estimator we propose, we let the nodes run independently and only during the partial aggregation stage we combine the samples from all the nodes as $S$. Thus, nodes operate asynchronously at different speed and produce samples with different size. The global randomization guarantees though that the combined sample $S$ is indeed a sample over the entire dataset. As a result, the generic sampling estimator in Equation~(\ref{eq:gen-est}) can be directly applied to this distributed setting without any modifications.

\subsubsection{Multiple Estimators Sampling}\label{sssec:est-agg-more}

For multiple estimators, the aggregate $\sum_{d \in D, \texttt{P}(d)} \texttt{f}(d)$ can be decomposed as $\sum_{i = 1}^{N} \sum_{d \in D_{i}, \texttt{P}(d)} \texttt{f}(d)$, with each node computing the sum over the local partition in the first stage followed by summing-up the local results to get the overall result in the second stage. An estimator $X_{i} = \frac{|D_{i}|}{|S_{i}|} \sum_{s \in S_{i}, \texttt{P}(s)} \texttt{f}(s)$ is defined for each partition based on the generic sampling estimator in Equation~(\ref{eq:gen-est}). We can then immediately infer that the sum of the estimators $X_{i}$, $\sum_{i = 1}^{N} X_{i}$, is an unbiased estimator for the query result and derive the variance $\Var{\sum_{i = 1}^{N} X_{i}} = \sum_{i = 1}^{N} \Var{X_{i}}$ if the sampling process across partitions is independent. Since the samples are taken independently from each data partition, local randomization of the data at each processing node is sufficient for the analysis to hold.

\subsubsection{Discussion}\label{sssec:est:discussion}

We propose an estimator for parallel online aggregation based on the \textit{single estimator} approach. The main difference is that our estimator is completely asynchronous and allows fully parallel evaluation. We show how it can be derived and analyzed starting from a generic sampling estimator for centralized settings. The implementation in PF-OLA for three different aggregation problems of various complexity is presented in Section~\ref{sec:est-examples}. We conclude with a detailed comparison with a stratified sampling estimator (or \textit{multiple estimators}) along multiple dimensions:
\begin{compactitem}
\item \textit{Data randomization.} While the multiple estimators approach requires only local randomization, the single estimator approach requires global randomization across all the nodes in the system. Although this might seem a demanding requirement, the randomization process can be entirely overlapped with data loading as part of hash-based data partitioning.
\item \textit{Dataset information.} Multiple estimators requires each node to have knowledge of the local partition cardinality, i.e., $|D_{i}|$. Single estimator needs only full cardinality information, i.e., $|D|$, where the estimation is invoked.
\item \textit{Accuracy.} According to the stratified sampling theory, multiple estimators provides better accuracy when the size of the sample at each node is proportional with the local dataset size (but not a requirement)~\cite{sampling-techniques}. This is not true in the general case though with the variance of the estimators being entirely determined by the samples at hand. Given that PF-OLA is a highly asynchronous framework, this optimal condition is hard to enforce.
\item \textit{Convergence rate.} As with accuracy, it is not possible to characterize the relative convergence rate of the two methods in the general case. Nonetheless, we can argue that multiple estimators is more sensitive to discrepancies in processing across the nodes since the effect on variance is only local. Consider for example the case when one variance is considerably smaller than the others. Its effect on the overall variance is asymptotically limited by the fraction it represents from the overall variance rather than the overall variance.
\item \textit{Fault tolerance.} The effect of node failure on the estimation process is catastrophic for multiple estimators. If one node cannot be accessed, it is impossible to compute the estimator and provide bounds since the corresponding variance is infinite. For single estimator, the variance decrease stops at a higher value than zero. This results in bounds that do not collapse on the true result even when all the available data is processed. 
\end{compactitem}

\section{PF-OLA Framework Design}\label{sec:pf-ola}

Online aggregation in PF-OLA is a process consisting of multiple stages. In the first stage, data are stored in random order on disk. This is an offline process executed at load time. Once data are available, we can start executing aggregate queries. A query specifies the input data and the computation to be executed. Every computation, including estimation, is expressed as a UDA---the central abstraction in the PF-OLA framework. The user application submits the query to a coordinator---a designated node managing the execution. It is the job of the coordinator to send the query further to the worker nodes, coordinate the execution, and return the result to the user application once the query is done. The result of a query is always a UDA containing the final state of the aggregate---PF-OLA takes UDAs as input and produces a UDA as output by implementing the UDA evaluation mechanism. When executing the query in non-interactive mode, the user application blocks until the final UDA arrives from the coordinator. When running in interactive mode with online aggregation enabled, the user application emits requests to the coordinator asking for the current state of the computation. Evidently, the returned UDA has an incomplete state since it is computed only over a portion of the data. The user application can use this partial UDA in many different ways. Computing online estimators and confidence bounds is only one of them.

To our knowledge, PF-OLA is the first online aggregation system in which the estimation is driven by the user application rather than the system. The main reason for this design choice is our goal to allow maximum asynchrony between the nodes in the system and to minimize the number of communication messages. A query always starts executing in non-interactive mode. Partial results are extracted asynchronously based only on user application requests. It is clear that generating partial results interferes with the actual computation of the query result. In the case of aggregate computation though, this is equivalent to early aggregation which is executed as part of the final aggregation nonetheless. Given this high-level description of the framework, we concentrate our attention on the following two important questions:
\begin{itemize}
\item How to enhance the UDA interface with estimation functionality without modifying the execution model?
\item How to optimally overlap the estimation process with query execution?
\end{itemize}

\subsection{User-Defined Aggregates (UDA)}\label{ssec:pf-ola:uda}

\begin{figure}
\centering
\includegraphics[width=0.85\textwidth]{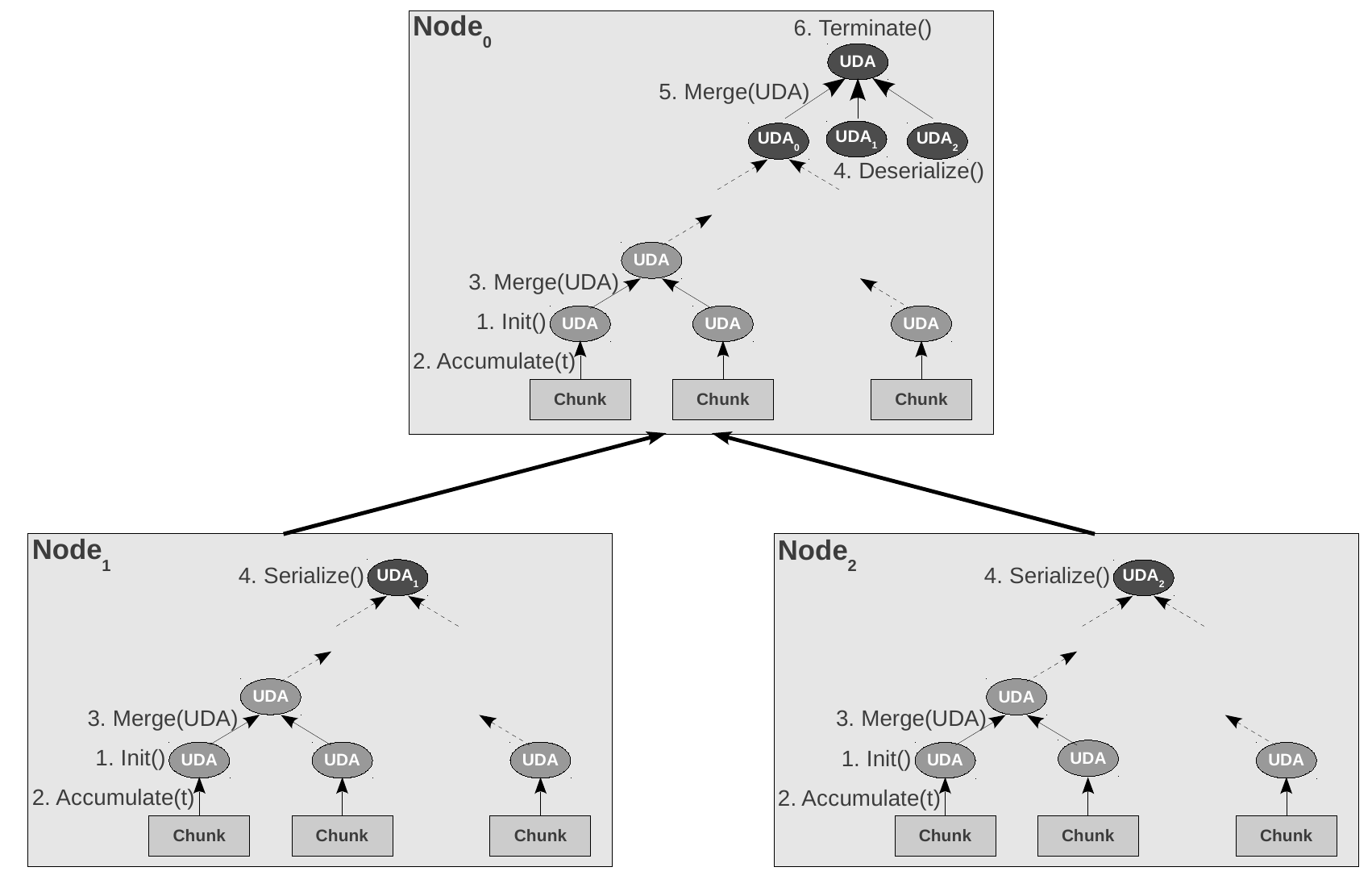}
\caption{UDA evaluation mechanism.}
\label{fig:UDA}
\end{figure}

UDAs~\cite{uda} represent a mechanism to extend the functionality of a database with application-specific aggregate operators similar in nature to user-defined data types (UDT) and user-defined functions (UDF)~\cite{new-types-in-POSTGRES}. A UDA is typically implemented as a class with a standard interface defining the following four methods~\cite{uda:theory}: \texttt{Init}, \texttt{Accumulate}, \texttt{Merge}, and \texttt{Terminate}. These methods operate on the \texttt{state} of the aggregate which is also part of the class. While the interface is standard, the user has complete freedom when defining the \texttt{state} and implementing the methods. The execution engine (runtime) computes the aggregate by scanning the input relation and calling the interface methods as follows. \texttt{Init} is called to initialize the state before the actual computation starts. \texttt{Accumulate} takes as input a tuple from the input relation and updates the state of the aggregate according to the user-defined code. \texttt{Terminate} is called after all the tuples are processed in order to finalize the computation of the aggregate. \texttt{Merge} is not part of the original specification~\cite{uda} and is intended for use when the input relation is partitioned and multiple UDAs are used to compute the aggregate (one for each partition). It takes as parameters two UDAs and it merges their states into the state of an output UDA. In the end, all the UDAs are merged into a single one upon which \texttt{Terminate} is called. A graphical depiction of the entire execution process is shown in Figure~\ref{fig:UDA}.

In order to provide online estimates, it is clear that additional data need to be stored in the UDA \texttt{state}. The exact data are determined by the actual estimator. Since users implement the UDA, they have complete control over what data go in the \texttt{state}. What users do not have control over though is the estimation process, driven by the UDA interface and the invocation mechanism. This imposes restrictions on the set of estimators that can be implemented using the standard UDA interface. Thus, the UDA interface also needs to be extended with additional functions. Our goal is to limit the modifications to the UDA interface and to the overall invocation process while still allowing for any estimator to be implemented in the PF-OLA framework.

One of the main contributions made in this paper is the design of extensions to the UDA interface for online aggregation and the subsequent implementation in PF-OLA. The first extension handles the communication problem. Since UDAs are transferred between nodes and from the coordinator to the user application, a mechanism that does this transparently is required. Given that the UDA state is defined by the user, it is the writer of the UDA who is in the position to specify what needs to be transferred in order to re-create an equivalent UDA in the memory space of another process. Thus, it is natural to apply the same principles at the core of UDA and extend the UDA interface with methods to \texttt{Serialize/Deserialize} the UDA state. It is the job of the UDA creator to implement these methods correctly and the responsibility of the framework to invoke them transparently.

The second extension is specifically targeted at estimation modeling for online aggregation. To support estimation, the UDA state needs to be enriched with additional data on top of the original aggregate. Although it is desirable to have a perfect overlap between the final result computation and estimation, this is typically not possible. In the few situations when it is possible, no additional changes to the UDA interface are required. For the majority of the cases though, the UDA interface needs to be extended in order to distinguish between the final result and a partial result used for estimation. As we shall see in Section~\ref{sec:est-examples}, there are at least two methods that need to be added: \texttt{EstimatorTerminate} and \texttt{EstimatorMerge}. \texttt{EstimatorTerminate} computes a local estimator at each node. It is invoked after merging the local UDAs during the estimation process. \texttt{EstimatorMerge} is called to put together in a single UDA the estimators computed at each node by \texttt{EstimatorTerminate}. It is invoked with UDAs originating at different nodes. It is important to notice that \texttt{EstimatorTerminate} is an intra-node method while \texttt{EstimatorMerge} is inter-node. It is possible to further separate the estimation from aggregate computation and have an intra-node \texttt{EstimatorMerge} and an inter-node \texttt{EstimatorTerminate}. While this adds more flexibility and might be required for particular estimation models, we have not encountered such a situation.

The third extension we add to the UDA interface is the \texttt{Estimate} method. It is invoked by the user application on the UDA returned by the framework as a result of an estimation request. The complexity of this method can range from printing the UDA state to complex statistical models. In the case of online aggregation, \texttt{Estimate} computes an estimator for the aggregate result and corresponding confidence bounds. As with the other methods, the only restriction on \texttt{Estimate} is that it can only access the UDA state.

\begin{small}
\begin{table}[htbp]
    \begin{center}
      \begin{tabular}{|p{0.68\textwidth}||p{0.23\textwidth}|}

    \hline
	\textbf{Method} & \textbf{Usage} \tabularnewline

	\hline\hline
	\texttt{Init ()} & Basic interface \tabularnewline
	\texttt{Accumulate (Item $d$)} & \tabularnewline
	\texttt{Merge (UDA $input_{1}$,UDA $input_{2}$, UDA $output$)} & \tabularnewline 
	\texttt{Terminate ()} & \tabularnewline

	\hline
	\texttt{Serialize ()} & Transfer UDA \tabularnewline
	\texttt{Deserialize ()} & across processes \tabularnewline
	
	\hline
	\texttt{EstimatorTerminate ()} & Partial aggregate \tabularnewline
	\texttt{EstimatorMerge (UDA $input_{1}$,UDA $input_{2}$, UDA $output$)} & computation \tabularnewline
	
	\hline
	\texttt{Estimate ($estimator$, $lower$, $upper$, $confidence$)} & Online estimation\tabularnewline
	
	\hline

      \end{tabular}
    \end{center}

    \caption{Extended UDA interface.}
    \label{tbl:uda-interface}
\end{table}
\end{small}

Table~\ref{tbl:uda-interface} summarizes the extended UDA interface we propose for parallel online aggregation. This interface abstracts both the aggregation and estimation processes in a reduced number of methods, releasing the user from the details of the actual execution in a parallel environment which are taken care of transparently by PF-OLA. Thus, the user can focus only on estimation modeling. In the rest of the paper we assume this interface for all the UDA examples we provide. Whenever a method is missing, it is assumed that it does not change the UDA state. In~\cite{uda} the authors observed that the basic UDA interface could be applied to online aggregation. As we shall see in Section~\ref{sec:est-examples}, this is the case only for a reduced class of estimators.

\subsection{GLADE}\label{ssec:pf-ola:glade}

GLADE~\cite{GLADE-OSR,GLADE-Demo} is a parallel processing system optimized for the execution of Generalized Linear Aggregates (GLA). A GLA is an associative-decomposable UDA. Essentially, what this means is that the order in which \texttt{Accumulate} and \texttt{Merge} are invoked does not change the final result. This in turn allows for efficient asynchronous computation in a parallel environment along an aggregation tree (Figure~\ref{fig:UDA}). The UDAs that do not satisfy the associative-decomposable condition require strict ordering when merged together, typically enforced through synchronization at a single node in the system---all the UDAs are first gathered at the node and then \texttt{Merge} is called in the proper order. GLADE can also handle this more general type of UDA at the expense of a considerable drop in performance. Since the online aggregation formulation (Equation~\ref{eq:prob-def}) we consider in this paper is associative-decomposable, thus a GLA, we choose GLADE as the runtime environment of our PF-OLA framework. While GLADE can execute GLAs specified using the basic UDA interface, it cannot provide the partial aggregate state required for online estimation. In this section, we show how to extend GLADE in order to handle GLAs represented through the extended UDA interface. This is quite a challenging task when both execution efficiency and estimation convergence have to be optimized---the case in overlapped online aggregation.

To understand how online aggregation is implemented in GLADE, we first show how GLA processing is realized. As with any other parallel system, GLADE consists of two types of nodes---one coordinator and multiple worker nodes. All GLA computations are directed to the coordinator who is responsible for setting-up the data flow between nodes -- building the aggregation tree -- and managing the execution. The coordinator compiles the GLA code based on the query parameters and then ships it together with a job description to the nodes. The worker nodes have both processing resources as well as attached storage. They act as independent entities, each running a GLA-enhanced instance of the DataPath~\cite{datapath} database system. The node loads the compiled GLA code inside DataPath and then executes it on the local data. As a final step, the resulting GLAs are put together along the tree structure using the \texttt{Merge} method in the UDA interface. While all this seems standard parallel database processing, adding GLAs as first class citizens inside a relational database such as DataPath is far from trivial.

\subsection{DataPath}\label{ssec:pf-ola:datapath}

DataPath~\cite{datapath} is a centralized multi-query processing database for analytics capable of providing single-query performance even when a large number of (different) queries are running in parallel. The most distinguishable features it brings are a push-based execution model and full data sharing at all stages during query processing. It is the speed at which data are read from disk that drives the entire execution. Data are read from disk asynchronously by multiple parallel threads and pushed into the execution engine for processing. DataPath operators process data at chunk granularity -- typically a few million tuples -- in highly optimized loops that scan the tuples and call operator-specific tasks. Data parallelism is achieved by processing multiple chunks in parallel---inside the same operator and across operators. Pipelining is implemented both at query plan level -- chunks are passed between operators -- as well as operator level---operators are decomposed into small tasks that are executed concurrently. The DataPath execution engine has at its core two key components---waypoints and work units. A waypoint manages a particular type of computation, e.g., selection, join, GLA. It is not executing any query processing by itself. It only delegates a particular task to a work unit. A work unit is a thread from a thread pool -- there are a fixed number of work units in the system -- that can be configured to execute tasks.

Intuitively, we can think of the waypoint as the control unit in a pipelined microprocessor where each work unit corresponds to a stage in the pipeline. Moreover, to support data parallelism, multiple such pipelines are created dynamically during query execution in what resembles a super-scalar microprocessor. For complete generalization, the same principle is applied at the level of each waypoint, resulting in multiple super-scalar units with different structure and functionality. Unlike in hardware where the configuration is static and it cannot be modified, the DataPath configuration is dynamic even for a given processing task. It adapts dynamically to the actual data flow generated from disk, with threads being assigned to the work units that require processing. If no threads are available, the chunk is dropped, all the computation up to that point is lost, and it has to be re-read from disk. As long as the computation is I/O-bound -- the case in databases -- this should not happen (at all) frequently.

The entire GLA processing is embedded inside the \texttt{GLAWaypoint} which consists of a pipeline of two work units -- one corresponding to \texttt{Accumulate} and one to \texttt{Merge} -- that invoke the corresponding methods in the UDA interface. The \texttt{GLAWaypoint} is not aware of the exact type of GLA it is executing since all it has to do is to relay chunks and tasks to work units that execute the actual work. It needs to store though the state of the GLA it is computing. More precisely, a list of GLA states is stored to allow multiple chunks to be processed in parallel. Then, the computation of a GLA proceeds as follows:
\begin{compactenum}
\item When a chunk needs to be processed, the \texttt{GLAWaypoint} extracts a GLA state from the list and passes it together with the chunk to a work unit (WorkUnit$_{1}$ in Figure~\ref{ssec:pf-ola:fig}). The task executed by the work unit is to call \texttt{Accumulate} for each tuple in the chunk such that the GLA is updated with all the tuples in the chunk. If no GLA state is passed with the task, a new GLA is created and initialized (\texttt{Init}) inside the task, such that a GLA is always sent back to the \texttt{GLAWaypoint}.
\item When all the chunks are processed, the list of GLA states has to be merged. Notice that the maximum number of GLA states that can be created is bounded by the number of work units in the system. The merging of two GLAs is done by another task that calls \texttt{Merge} on the two states (WorkUnit$_{2}$ in Figure~\ref{ssec:pf-ola:fig}). 
\item In the end, \texttt{Terminate} is called on the last state inside another task submitted to a work unit.
\end{compactenum}

\subsection{Parallel Online Aggregation in GLADE}\label{ssec:pf-ola:gla-w-ola}

\begin{figure}
\centering
\includegraphics[width=0.6\textwidth]{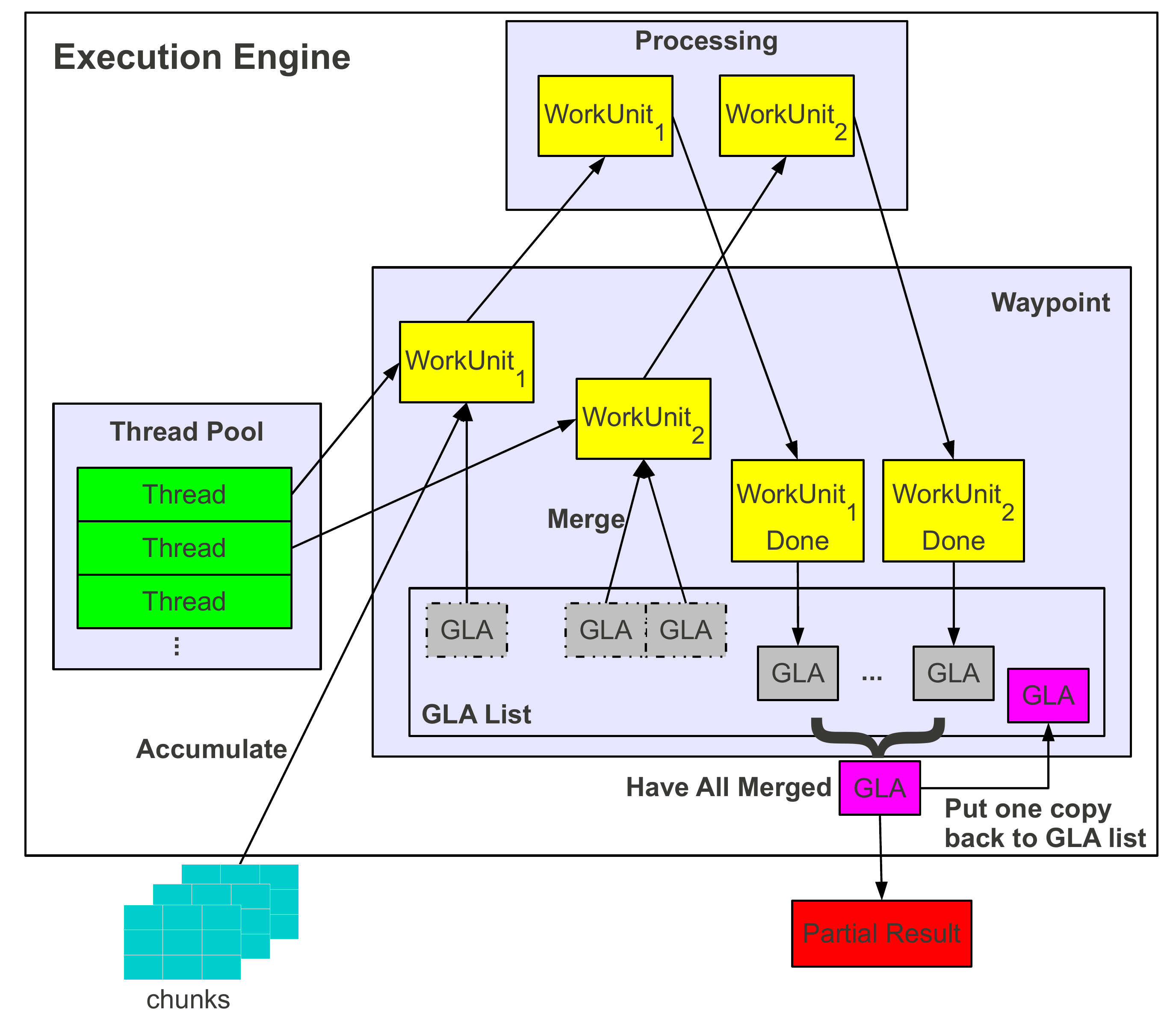}
\caption{Execution strategy for parallel online aggregation in PF-OLA.}
\label{ssec:pf-ola:fig}
\end{figure}

At a high level, enhancing GLADE with online aggregation is just a matter of providing support for GLAs expressed using the extended UDA interface in Table~\ref{tbl:uda-interface}. Intuitively, this can be done using the same mechanism we currently have in place inside the \texttt{GLAWaypoint}. While this is a good starting point, there are multiple aspects that require careful consideration. For instance, the system is expected to process partial result requests at any rate, at any point during query execution, and with the least amount of synchronization among the processing nodes. Moreover, the system should not incur any overhead on top of the normal execution when online aggregation is enabled. Under these requirements, the task becomes quite challenging. Our solution overlaps online estimation and actual query processing at all levels of the system and applies multiple optimizations.

We present how online aggregation requests are processed by following the path of the message in the system. In the normal situation, a partial result request follows exactly the same path as a new GLA computation. Problems arise when multiple requests overlap, when requests are sent during the final result aggregation phase, and when there are nodes who have finished local processing and nodes who have not. All these scenarios are likely to appear given the asynchronous nature of the requests and of the processing. Our solution handles each of these situations as follows:
\begin{compactitem}
\item A partial result request which arrives while a particular node is still processing a previous request is discarded. The partial result generated for the earliest request is returned for all the discarded requests.
\item A partial request received at the DataPath execution engine during the local GLA aggregation stage -- final result aggregation stage -- is discarded given that the same final GLA would be produced nonetheless.
\item If a node has finished local processing and a partial result request is received, the node returns the local GLA for merging with the partial/final GLAs from the other nodes. This allows for online estimates to be produced even when there is considerable discrepancy in processing speed across nodes.
\end{compactitem}

Once the request arrives at the \texttt{GLAWaypoint}, it is not clear how to proceed. The obvious choice is to process exclusively the partial result computation in order to generate the estimate and the bounds as soon as possible. This results in dropped chunks and thus incurs delays in the final GLA processing. It is though the only available choice in all online aggregation systems we are aware of. The solution we adopt in PF-OLA is to overlap online estimation and GLA processing. Abstractly, this corresponds to executing two simultaneous GLA computations. Given that DataPath is a multi-query processing system optimized for sharing data across multiple queries, optimal performance is expected. To get this performance though, careful consideration of several aspects is required. Rather than treating actual computation and estimation as two separate GLAs, we group everything into a single GLA satisfying the extended UDA interface (Table~\ref{tbl:uda-interface}). This simplifies the logic inside the \texttt{GLAWaypoint} to the following steps:
\begin{compactitem}
\item A partial result request triggers the merging of all existent GLAs, those inside the waypoint as well as those modified upon by a work unit. The resulting GLA is the partial result. Unlike the final result which is extracted from the waypoint and passed for further merging across the nodes, a copy of the partial result needs to be kept inside the waypoint and used for the final result computation (Figure~\ref{ssec:pf-ola:fig}).
\item The newly arriving chunks are processed as before. The result is a completely new list of GLA states. The local GLA resulted through merging is added to this new list once the merging process ends.
\end{compactitem}

To summarize, adding online aggregation to GLADE requires the extraction of a snapshot of the system state that can be used for estimation. Our solution overlaps the process of taking the snapshot with the actual GLA processing in order to have minimum impact on the overall execution time. This is facilitated to some extent by the multi-query processing capabilities of the underlying DataPath system.

\section{PF-OLA Estimation Examples}\label{sec:est-examples}

\begin{algorithm}[htbp]
\caption{\textit{GLASum-SingleEstimator}}
\label{alg:gla-sum-single}
\algsetup{linenodelimiter=.}

\textbf{State:} $\textit{sum}$; $\textit{sumSq}$; $\textit{count}$

\textbf{Init} ()
\begin{algorithmic}[1]
\STATE $\textit{sum} = 0$; $\textit{sumSq} = 0$; $\textit{count} = 0$
\end{algorithmic}

\textbf{Accumulate} (Tuple \texttt{t})
\begin{algorithmic}[1]
\IF {$\texttt{P}(\texttt{t})$}
\STATE $\textit{sum} = \textit{sum} + \texttt{f}(\texttt{t})$; $\textit{sumSq} = \textit{sumSq} + \texttt{f}^{2}(\texttt{t})$; $\textit{count} = \textit{count} + 1$
\ENDIF
\end{algorithmic}

\textbf{Merge} (GLASum\hspace*{0.09cm}$input_{1}$,\hspace*{0.09cm}GLASum\hspace*{0.09cm}$input_{2}$,\hspace*{0.09cm}GLASum\hspace*{0.09cm}$output$)\hspace*{0cm}
\begin{algorithmic}[1]
\STATE $output.sum = input_{1}.sum + input_{2}.sum$
\STATE $output.sumSq = input_{1}.sumSq + input_{2}.sumSq$
\STATE $output.count = input_{1}.count + input_{2}.count$
\end{algorithmic}

\textbf{Terminate} ()

\textbf{Estimate} ($\textit{estimator}$,\hspace*{0.12cm}$\textit{lowerBound}$,\hspace*{0.12cm}$\textit{upperBound}$,\hspace*{0.12cm}$\textit{confLevel}$)\hspace*{0.12cm}
\begin{algorithmic}[1]
\STATE $estimator = \frac{|D|}{count} * sum$
\STATE $estVar = \frac{|D|*(|D|-count)}{count^{2}*(count-1)} * \left(count * sumSq - sum^{2} \right)$
\STATE $lowerBound = estimator + NormalCDF \left( \frac{1 - \textit{confLevel}}{2}, \sqrt{estVar} \right)$
\STATE $upperBound = estimator + NormalCDF \left( \textit{confLevel} + \frac{1 - \textit{confLevel}}{2}, \sqrt{estVar} \right)$
\end{algorithmic}
\end{algorithm}

In this section, we show how three different online estimation problems of complexity ranging from simple and \texttt{GROUP BY} aggregation to general SPJ queries are expressed in the PF-OLA framework. We present the GLAs for single and multiple estimators, respectively, and discuss the most significant implementation aspects.

\subsection{Aggregation}\label{ssec:est:agg}

The first online aggregation problem we consider is single-table aggregation. Specifically, we consider aggregates of the following SQL form:\\
\begin{equation}\label{eq:single-table-SQL}
\begin{split}
&\texttt{SELECT SUM(f(t))}\\
&\texttt{FROM TABLE AS t}\\
&\texttt{WHERE P(t)}
\end{split}
\end{equation}
which compute the SUM of function \texttt{f} applied to each tuple in table \texttt{TABLE} that satisfies condition \texttt{P}. It is straightforward to express the aggregate~(\ref{eq:single-table-SQL}) in the GLA form since it is an associative-decomposable expression. The \texttt{state} consists only of the running sum, initialized at zero. \texttt{Accumulate} updates the current sum with \texttt{f(t)} only for the tuples \texttt{t} satisfying the condition \texttt{P}, while \texttt{Merge} adds the states of the input GLAs and stores the result as the state of the output GLA.

Algorithm~\textit{GLASum-SingleEstimator} implements the estimator we propose. No modifications to the UDA interface are required. Looking at the GLA state, it might appear erroneous that no sample is part of the state when a sample over the entire dataset is required in the estimator definition. Fortunately, the estimator expectation and variance can be derived from the three variables in the state computed locally at each node and then merged together globally. This reduces dramatically the amount of data that needs to be transferred between nodes. To compute the estimate and the bounds, knowledge of the full dataset size is required in \texttt{Estimate}.

\begin{algorithm}[htbp]
\caption{\textit{GLASum-MultipleEstimators}}
\label{alg:gla-sum-mult}
\algsetup{linenodelimiter=.}

\textbf{State:} $\textit{sum}$; $\textit{sumSq}$; $\textit{count}$; $\textit{est}$; $\textit{estVar}$

\textbf{EstimatorTerminate} ()
\begin{algorithmic}[1]
\STATE $est = \frac{|D_{i}|}{count} * sum$
\STATE $estVar = \frac{|D_{i}|*(|D_{i}|-count)}{count^{2}*(count-1)} * \left(count * sumSq - sum^{2} \right)$
\end{algorithmic}

\textbf{EstimatorMerge} (GLASum $input_{1}$, GLASum $input_{2}$, GLASum $output$)
\begin{algorithmic}[1]
\STATE $output.est = input_{1}.est + input_{2}.est$
\STATE $output.estVar$\hspace*{0.02cm}$=$\hspace*{0.02cm}$input_{1}.estVar+$\hspace*{0.02cm}$input_{2}.estVar$\hspace*{0cm}
\end{algorithmic}

\textbf{Estimate} ($\textit{estimator}$,\hspace*{0.1cm}$\textit{lowerBound}$,\hspace*{0.1cm}$\textit{upperBound}$,\hspace*{0.1cm}$\textit{confLevel}$)\hspace*{0cm}
\begin{algorithmic}[1]
\STATE $estimator = est$
\STATE $lowerBound = estimator + NormalCDF \left( \frac{1 - \textit{confLevel}}{2}, \sqrt{estVar} \right)$
\STATE $upperBound = estimator + NormalCDF \left( \textit{confLevel} + \frac{1 - \textit{confLevel}}{2}, \sqrt{estVar} \right)$
\end{algorithmic}

\end{algorithm}

Algorithm~\textit{GLASum-MultipleEstimators} contains the GLA implementation for the multiple estimators approach. The extended UDA interface has to be used in this case. The standard UDA interface methods are identical to Algorithm~\textit{GLASum-SingleEstimator} for single estimator and are not repeated. The \texttt{state} consists of variables to compute the query result and local estimators -- $\textit{sum}$, $\textit{sumSq}$, $\textit{count}$ -- and variables to compute the global estimator---$\textit{est}$, $\textit{estVar}$. This separation is necessary since the estimation process and the query result computation are not entirely overlapped. The only difference in \texttt{Estimate} when compared to the single estimator approach is the scaling factor compensating for the local dataset size rather than the overall size. Method \texttt{EstimatorTerminate} is executed at each node to compute the local estimator corresponding to an estimation stage once the local merging finished. The two local estimators $\textit{est}$ and $\textit{estVar}$ are then merged across all the processing nodes in \texttt{EstimatorMerge} to complete the estimation stage. \texttt{Estimate} is then called on the resulting GLA to get the estimator and bounds.

\subsection{Group-By Aggregation}\label{ssec:est:group-agg}

\begin{algorithm}[htbp]
\caption{\textit{GLAGroupBy}}
\label{alg:gla-group-by}
\algsetup{linenodelimiter=.}

\textbf{State:} $\textit{groups} : \textit{Map(\texttt{gAtts}} \mapsto \textit{GLASum-SingleEstimator})$

\textbf{Init} ()

\textbf{Accumulate} (Tuple \texttt{t})
\begin{algorithmic}[1]
\IF {\texttt{t.gAtts} is not in \textit{groups}}
\STATE Insert \texttt{t.gAtts} into \textit{groups}
\STATE $\textit{groups}[\texttt{t.gAtts}].Init()$
\ENDIF
\STATE $\textit{groups}[\texttt{t.gAtts}].Accumulate(\texttt{t})$
\end{algorithmic}

\textbf{Merge} (GLAGroupBy $input_{1}$, GLAGroupBy $input_{2}$, GLAGroupBy $output$)
\begin{algorithmic}[1]
\STATE $output.groups = input_{1}.groups$
\FORALL {groups $g \in input_{2}.groups$}
\IF {$g.\texttt{gAtts}$ is in \textit{output.groups}}
\STATE Merge $\textit{g.GLA}$ into $\textit{output.groups}[g.\texttt{gAtts}]$
\ELSE
\STATE Insert $g$ into \textit{output.groups}
\ENDIF
\ENDFOR
\end{algorithmic}

\textbf{Terminate} ()

\textbf{Estimate} ($\textit{estimator}$, $\textit{lowerBound}$, $\textit{upperBound}$,$\textit{confLevel}$, $\textit{group}$)
\begin{algorithmic}[1]
\STATE $\textit{groups}[\textit{group}].\textit{Estimate}(\textit{estimator}, \textit{lowerBound}, \textit{upperBound}, \textit{confLevel})$
\end{algorithmic}

\end{algorithm}

The second problem we address is \texttt{GROUP BY} aggregation. Specifically, we consider queries of the following SQL form:
\begin{equation}\label{eq:group-by-SQL}
\begin{split}
&\texttt{SELECT gAtts, SUM(f(t))}\\
&\texttt{FROM TABLE AS t}\\
&\texttt{WHERE P(t)}\\
&\texttt{GROUP BY gAtts}
\end{split}
\end{equation}
which compute the same type of aggregate functions as (\ref{eq:single-table-SQL}) over the partitions induced by the grouping attributes \texttt{gAtts} rather than over the entire table. Notice that query (\ref{eq:group-by-SQL}) can be rewritten as multiple queries of the form (\ref{eq:single-table-SQL}), one for each group, with each distinct value of the grouping attributes encoded as a condition in \texttt{P}. Based on this formulation, we define an independent estimator for each group and derive confidence bounds accordingly. As long as there is no correlation between the grouping attributes and the aggregate function, the analysis provided for the generic sampling estimator holds. The extension to the PF-OLA parallel setting follows the same principles of single and multiple estimators, respectively, applied to individual groups. The reason we treat \texttt{GROUP BY} aggregation separately is to illustrate GLA composition. Instead of having different entities for the \texttt{GROUP BY} operator and UDAs, \texttt{GROUP BY} is represented as a GLA with composite state containing GLAs for the aggregate function. 

Algorithm~\textit{GLAGroupBy} gives the implementation of the \texttt{GROUP BY} GLA for the single estimator approach. The GLA \texttt{state} consists of a hash table that maps the aggregate GLAs to the corresponding grouping attributes. For the purpose of the presentation, the selection predicate \texttt{P} is kept inside the aggregate GLA. The estimation logic is encapsulated in the aggregate GLA -- in this case \textit{GLASum-SingleEstimator} -- with the \textit{GLAGroupBy} acting as a wrapper that directs the method calls to the correct group (\texttt{Accumulate} and \texttt{Estimate}). \texttt{Merge} invokes merging at the aggregate GLA level whenever a group is present in both input arguments and copies the entries otherwise. Since the GLA for multiple estimators is similar, it is not included.

\subsection{Join Group-By Aggregation}\label{ssec:est:join-group-agg}

Following the trend of increasing complexity, we focus now on the general join \texttt{GROUP BY} aggregation problem having the SQL form:
\begin{equation}\label{eq:join-SQL}
\begin{split}
&\texttt{SELECT gAtts, SUM(f($\texttt{t}_{\texttt{1}}$ $\bullet$ $\texttt{t}_{\texttt{2}}$))}\\
&\texttt{FROM $\texttt{TABLE}_{\texttt{1}}$ AS $\texttt{t}_{\texttt{1}}$, $\texttt{TABLE}_{\texttt{2}}$ AS $\texttt{t}_{\texttt{2}}$}\\
&\texttt{WHERE P($\texttt{t}_{\texttt{1}}$ $\bullet$ $\texttt{t}_{\texttt{2}}$)}\\
&\texttt{GROUP BY gAtts}
\end{split}
\end{equation}

In order to execute such a query in a parallel database, it is evident that tuples with the same join attributes in the two relations need to reside on the same node. We restrict the query to cases where $\texttt{TABLE}_{\texttt{2}}$ is a replicated relation across all the nodes, small enough to fit in memory. This allows us to represent the computation as a composite GLA with $\texttt{TABLE}_{\texttt{2}}$ being part of the GLA \texttt{state} together with a \textit{GLAGroupBy} instance. It also simplifies the estimation logic, encapsulated entirely in the \textit{GLAGroupBy}. 

\begin{algorithm}[htbp]
\caption{\textit{GLAJoin}}
\label{alg:gla-join}
\algsetup{linenodelimiter=.}

\textbf{State:} $H : \textit{join hash table}$, $groups : \textit{GLAGroupBy}$

\textbf{Init} ()
\begin{algorithmic}[1]
\STATE Initialize join hash table $H$ with tuples in $\texttt{TABLE}_{\texttt{2}}$
\end{algorithmic}

\textbf{Accumulate} (Tuple $\texttt{t}_{\texttt{1}}$)
\begin{algorithmic}[1]
\FORALL {join tuples \texttt{t} = $\texttt{t}_{\texttt{1}}$ $\bullet$ $\texttt{t}_{\texttt{2}}$ generated by $\texttt{t}_{\texttt{1}}$ in $H$}
\STATE $\textit{groups}.Accumulate(\texttt{t})$
\ENDFOR
\end{algorithmic}

\textbf{Merge} (GLAJoin $input_{1}$, GLAJoin $input_{2}$, GLAJoin $output$)
\begin{algorithmic}[1]
\STATE $Merge(input_{1}.groups, input_{2}.groups, output.groups)$
\end{algorithmic}

\textbf{Terminate} ()

\textbf{Estimate} ($\textit{estimator}$, $\textit{lowerBound}$, $\textit{upperBound}$, $\textit{confLevel}$, $\textit{group}$)
\begin{algorithmic}[1]
\STATE $\textit{groups}.\textit{Estimate}(\textit{estimator}, \textit{lowerBound}, \textit{upperBound}, \textit{confLevel}, \textit{group})$
\end{algorithmic}

\end{algorithm}

Algorithm~\textit{GLAJoin} contains the implementation of the aggregate (\ref{eq:join-SQL}) as a GLA based on the traditional UDA interface. In \texttt{Init}, the hash table corresponding to $\texttt{TABLE}_{\texttt{2}}$ is built in memory. Once this blocking process is over, the items in $\texttt{TABLE}_{\texttt{1}}$ are processed by \texttt{Accumulate}. First, the join tuples are found in the hash table $H$ and each of them is passed into \textit{GLAGroupBy} for further accumulation. \texttt{Merge} and \texttt{Estimate} are wrapper calls to corresponding methods in \textit{GLAGroupBy}. Given that hash table $H$ is built during the initialization, it is clear that no estimates can be computed until this phase is over. This is perfectly acceptable if we consider initialization to be part of query setup rather than query execution. Since in the PF-OLA framework $H$ is built by the user application and passed for execution to the processing nodes together with the query, this is the case.

In the following, we discuss generalizations to cases when $\texttt{TABLE}_{\texttt{2}}$ is not replicated across all the processing nodes and when it is too large to fit in memory. The only solution for parallel join processing is to bring tuples having the same join key on the same node, i.e., partition the tables on the join key. This can be done offline, during data loading, or online, during query execution. Since offline partitioning is the standard procedure, we focus on this case. We argue that as long as the partitioning is based on a random hash function and all the partitions corresponding to one of the two relations fit in memory -- assume $\texttt{TABLE}_{\texttt{2}}$ for consistency -- the solution for the replicated case is immediately applicable. The only difference is that each local estimator corresponds to a partition rather than the entire domain. When not all partitions of any of the two relations fit in memory, extensions to the standard ripple join~\cite{ripple-join} such as SMS join~\cite{sms-join} or PR join~\cite{pr-join} have to be combined with data partitioning to design adequate estimators. The algorithm proposed in~\cite{scalable-hash-ripple} represents such an example. We do not provide details on such extensions since they are beyond the scope of this paper. Nonetheless, we plan to address this topic in future work.

\section{Empirical Evaluation}\label{sec:empirical}
%
%

Our experimental study evaluates the main characteristics of the PF-OLA framework:
\begin{compactitem}
\item The expressiveness to represent different estimation models using the extended UDA interface given in Table~\ref{tbl:uda-interface}. For this, we implement GLA instances corresponding to the three estimation models -- multiple estimators and single estimator with and without synchronization -- for each of the three problems discussed in Section~\ref{sec:est-examples} and check the ability of the framework to execute them.
\item The lack of noteworthy overhead incurred by the estimation process on top of the actual execution when the estimators are implemented and executed in PF-OLA. For this, we measure the execution time with and without estimation across multiple instances of the three aggregation problems with various degrees of complexity and across different system configurations.
\end{compactitem}

An equally important objective of the study is to provide a thorough comparison between the non-synchronized single estimator solution we propose in this paper and multiple estimators. To this end, we evaluate the ``time 'til utility'' (TTU)~\cite{turbo:dbo} or convergence rate of the estimators as a function of query selectivity, number of groups, and data skew; the correctness of the confidence bounds they produce; the overhead in execution time they incur; and their robustness in the face of worker node delays and failures. The effect of parallelism on the two estimators is also studied by measuring the scaleup across cluster instances with different number of nodes.

Although we also executed the same set of experiments for the synchronized version of the single estimator, we do not include the results in this comparison. The reason is that this estimation model incurs a factor of four delay in execution time and thus has a considerable higher TTU.

\begin{figure*}
\centering
\subfloat[]{\includegraphics[width=0.5\textwidth]{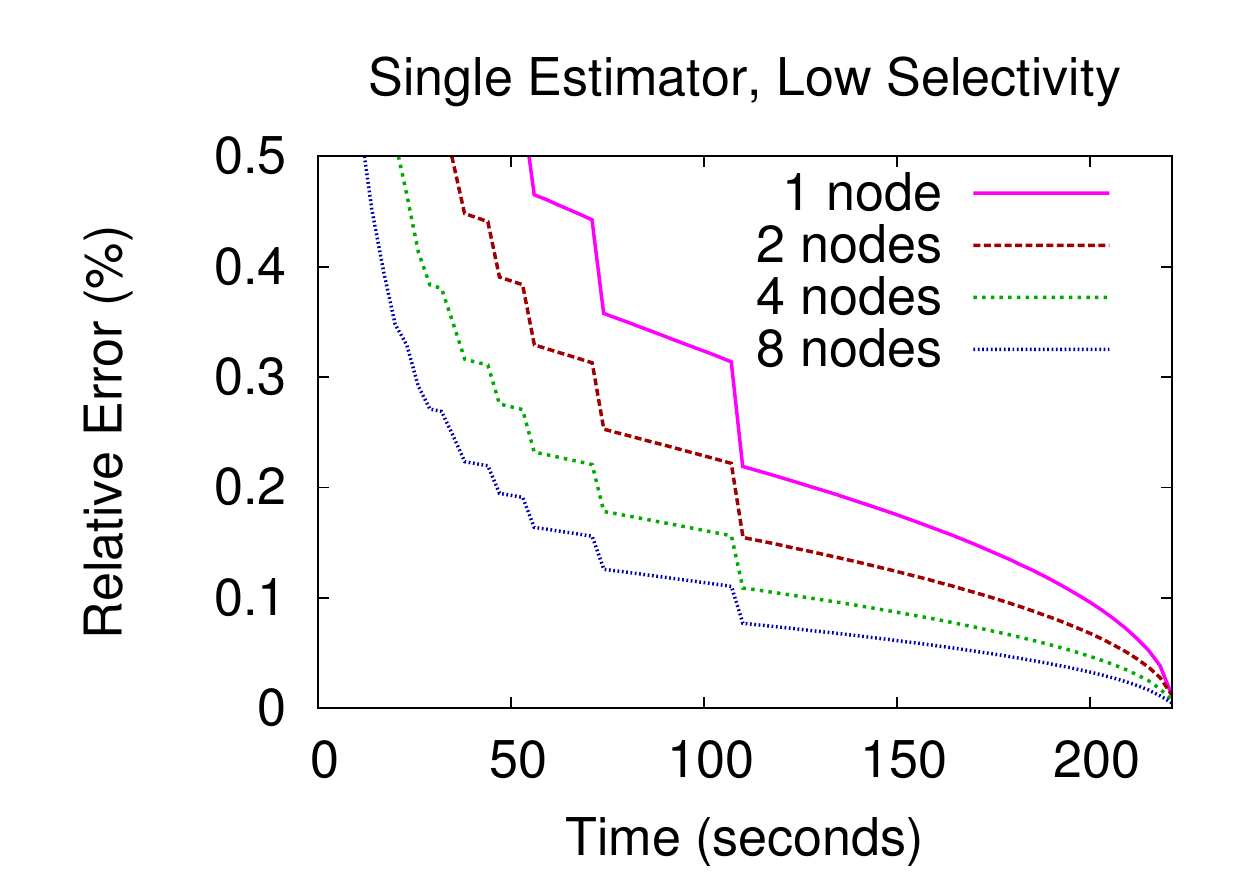}\label{fig:agg-single-low}}
\subfloat[]{\includegraphics[width=0.5\textwidth]{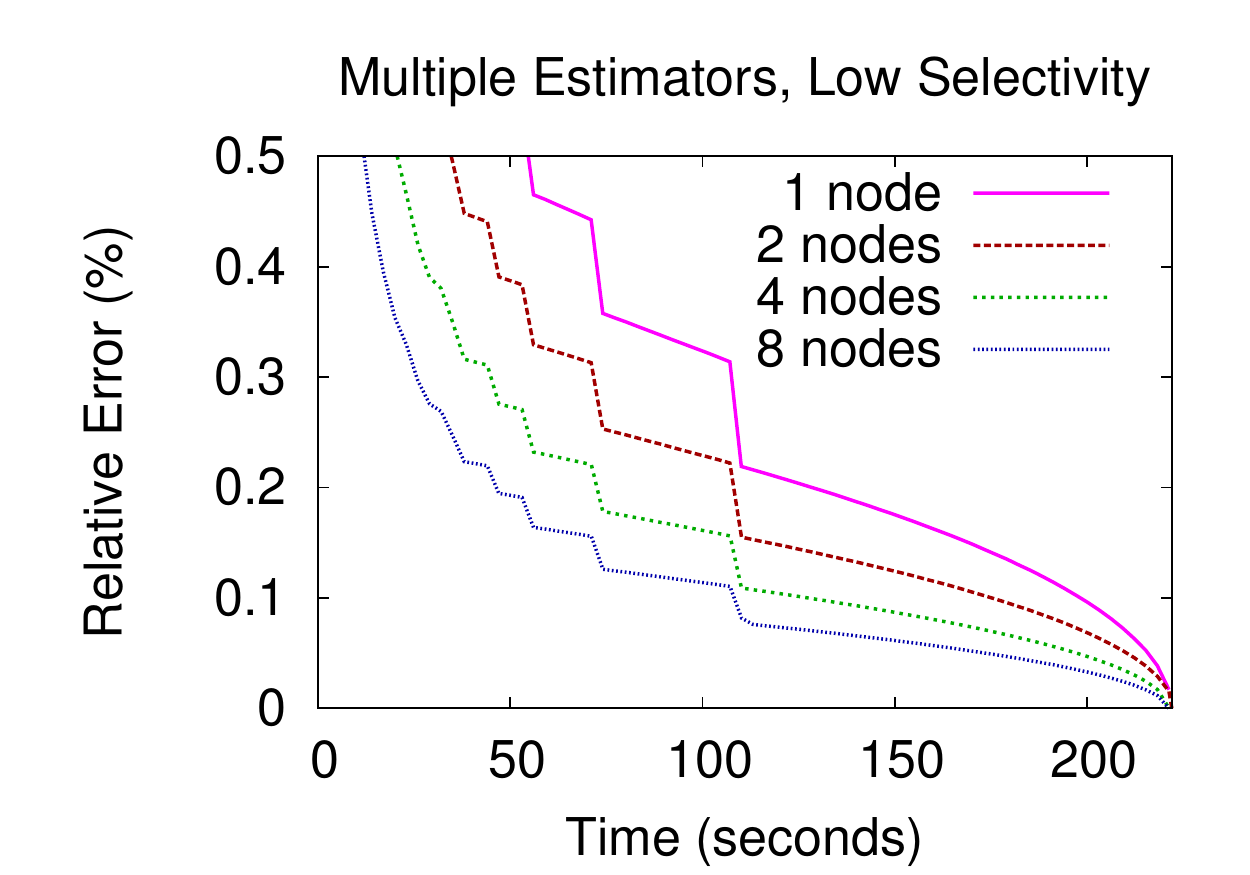}\label{fig:agg-multiple-low}}\\
\subfloat[]{\includegraphics[width=0.5\textwidth]{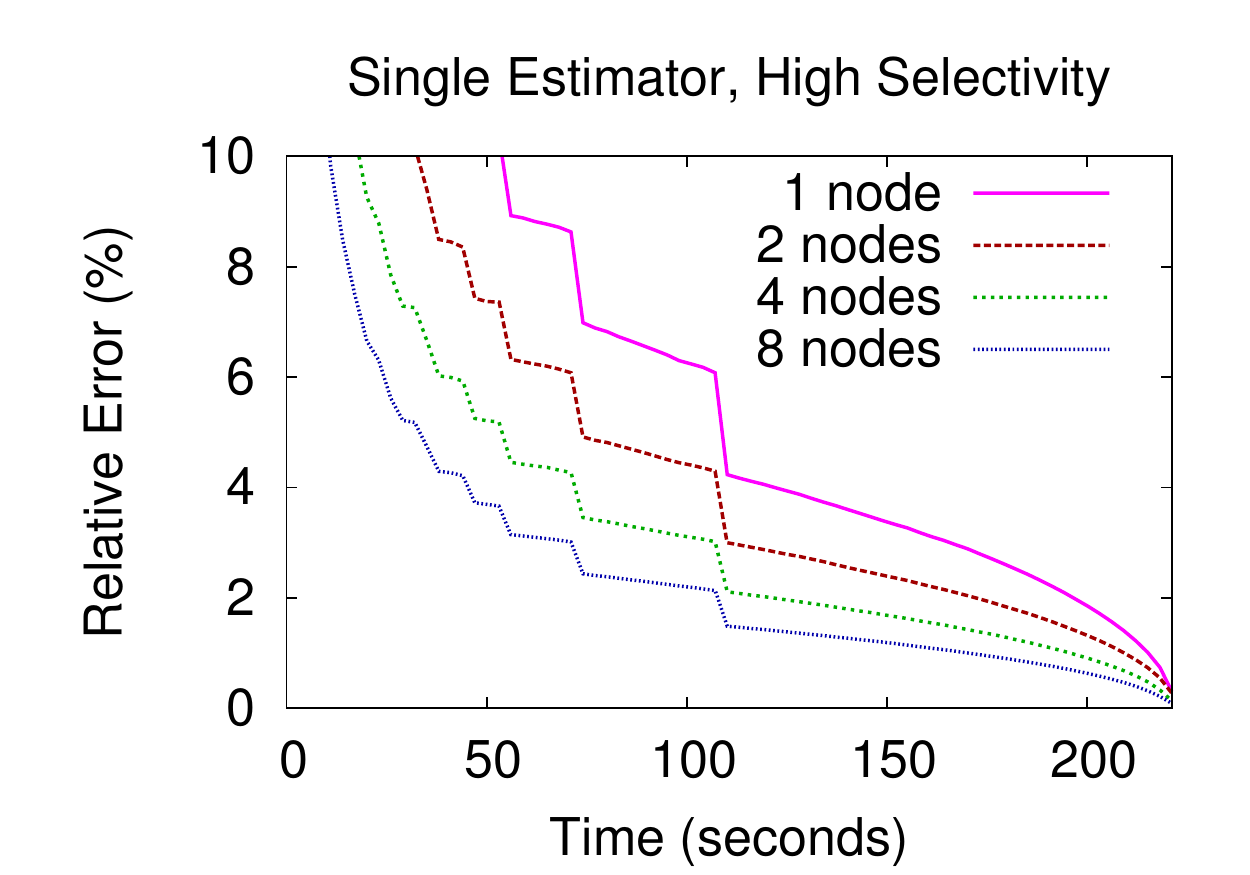}\label{fig:agg-single-high}}
\subfloat[]{\includegraphics[width=0.5\textwidth]{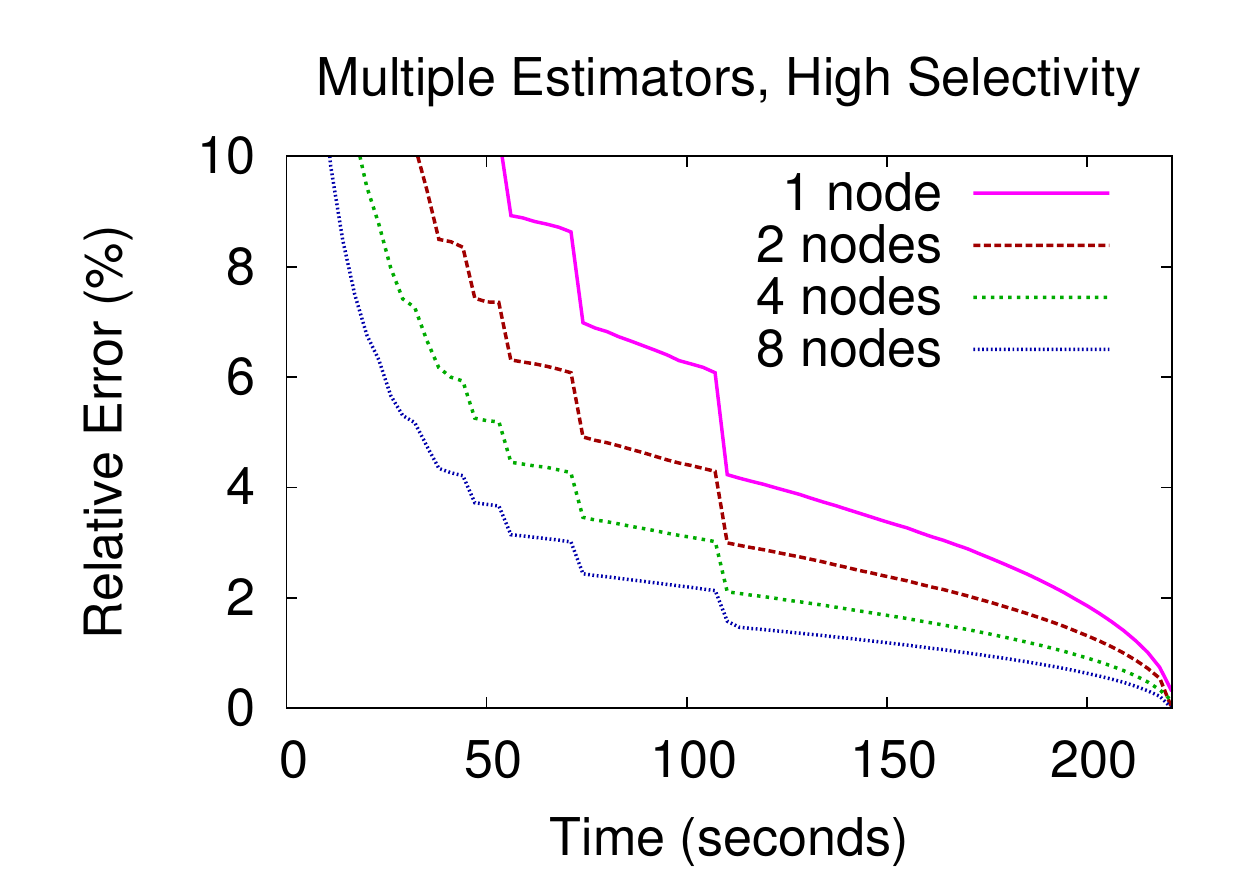}\label{fig:agg-multiple-high}}
\caption{Comparison between single estimator and multiple estimators for the aggregation task (TPC-H Q6). Figures (a) and (b) correspond to low selectivity (a large number of tuples satisfy the selection predicate), while Figures (c) and (d) correspond to high selectivity (only few tuples satisfy the selection condition).}
\label{fig:experiments:agg}
\end{figure*}

\subsection{Setup}\label{ssec:exp:setup}

\paragraph{Implementation.}
We implemented the PF-OLA framework as part of the GLADE parallel processing system~\cite{GLADE-Demo}. Since GLADE executes GLA computations specified using the basic UDA interface, we made considerable changes in order to enhance GLADE with support for the extended UDA interface. These changes involved modifications to the core execution engine, mainly the \texttt{GLAWaypoint}, as well as adding support for new types of messages at the communication infrastructure level. Overall, somewhere around $10,000$ lines of \texttt{C++} code. We emphasize that this is a one time effort to enhance the GLADE framework with online aggregation. Subsequently, the user has to implement only the extended GLA interface in order to design any estimation model---no changes to the system internals are required anymore, as is the case with all the previous online aggregation systems we are aware of. To prove our point, we successfully implemented all the proposed estimation models as GLAs. The result is a system that executes any GLA computation with or without online aggregation, depending on the runtime behavior of the user application.

All the GLAs presented in this study are implemented as standard \texttt{C++} classes providing the extended UDA interface and having the state represented as class member variables. The \texttt{template} mechanism and class composition are used extensively to avoid re-writing boiler-plate code for every single GLA. The actual computation is specified as a script containing the physical execution plan---input data sources and waypoints, the data flow between them, the association between GLAs and GLA waypoints, and the GLA arguments. Scripting provides a certain level of abstraction specific to a higher-level language and allows for significant code reuse.

\paragraph{System.}
We executed all our experiments on a $9$-node shared nothing cluster. Each node has $2$ AMD Opteron $8$-core processors for a total of $16$ cores running at $2\textit{GHz}$, $16\textit{GB}$ of memory, $4$ $1\textit{TB}$ hard-drives, and runs Ubuntu 10.10 $64$-bit. Out of the $16\textit{GB}$ memory, $12.5\textit{GB}$ are reserved as $6,000$ $2\textit{MB}$ huge pages for reduced TLB misses. The disks perform sequential reads at $130\textit{MB/s}$ according to \texttt{hdparm}, for $520\textit{MB/s}$ total I/O bandwidth at a node. The nodes are mounted inside the same rack and are inter-connected through a Gigabit Ethernet switch. One node is configured as the coordinator and the other $8$ nodes are workers.

\paragraph{Data.}
The dataset used in our experiments is TPC-H~\cite{tpch} scale $\textbf{8,000}$ ($\textbf{8\textit{TB}}$). Instead of generating the entire $8\textit{TB}$ at once (impossible due to the $4\textit{TB}$ limit on each node), we generated $8$ $1\textit{TB}$ independent TPC-H scale $1$ instances, one on every worker node. Notice that the instances are not copies of the same dataset and they are not treated as replicas---they are a single dataset partitioned into $8$ non-overlapping subsets. The data is then stripped across the $4$ available disks, for approximately $250\textit{GB}$ on each disk. Rather than executing a global data randomization process, we opted for local randomization due to efficiency reasons. This is supposed to have a negative effect on the accuracy and correctness of the estimator proposed in the paper since it relies on global randomization. Nonetheless, the experimental results -- the Monte Carlo simulations in particular -- prove this is not the case and show that local randomization is sufficient in the majority of the cases.

\paragraph{Methodology.}
All our figures depict relative confidence bounds width as a function of time for cluster configurations of $1$, $2$, $4$, and $8$ nodes, respectively. Relative confidence bounds width ($\frac{\textit{high} - \textit{low}}{\textit{estimate}}$) is measured as the ratio between the difference of the confidence interval extremes and the estimate. The bounds are computed at $95\%$ confidence level and are depicted as percentage from the estimate. The different cluster configurations test the scaleup of the PF-OLA framework and the effect of parallelism on the estimators. A system scales-up optimally if the execution time remains constant when both the data and the processing capacity increase proportionally, i.e., all curves collapse to zero at the same time in our figures.

\subsection{Aggregation}\label{ssec:exp:agg}

The aggregation tasks we consider are two different instantiations of query Q6 from the TPC-H benchmark:
\begin{equation*}\label{query:exp:agg}
\setlength{\jot}{0.1pt}
	\begin{split}
		& \texttt{SELECT SUM(l\_extendedprice*l\_discount)} \\
		& \texttt{FROM lineitem} \\
		& \texttt{WHERE l\_shipdate between [$\texttt{date}_{1}$, $\texttt{date}_{2}$] AND} \\
		& \hspace*{0.1cm}\texttt{l\_discount\hspace*{0.09cm}between\hspace*{0.09cm}[0.02, 0.03]\hspace*{0.09cm}AND\hspace*{0.09cm}\texttt{l\_quantity\hspace*{0.09cm}=\hspace*{0.09cm}1}} \\
	\end{split}
\end{equation*}
In the first instance, \texttt{l\_shipdate} takes values in the one year interval beginning on \texttt{`1993-02-26'}. This is considered a low selectivity query since a large number of tuples satisfy the selection predicate (approximately $13.3\times10^{6}$ out of $48\times10^{9}$ for the $8$ node configuration). The second instance restricts \texttt{l\_shipdate} to a single day, \texttt{`1993-02-26'}. It is a high selectivity query (or needle in the haystack) since only $35,000$ tuples are part of the result.

Figure~\ref{fig:experiments:agg} depicts the relative confidence bounds width as a function of query execution time for the two estimators considered in the study. There are multiple common trends across all the graphs. The width of the bounds decreases as the execution progresses and more tuples are processed, converging to the true result in the end. This is the standard online aggregation behavior. At the same instance in time, the width of the bounds is wider for the configurations with fewer nodes or, equivalently, it takes more time to arrive at the same relative width. This exemplifies perfectly the effect of parallelism on estimation. Having more nodes increases the rate at which result tuples are discovered thus reducing TTU. The execution time for all the configurations is the same. This confirms the scaleup of our framework. The confidence bounds are identical for the two estimators in the single node configuration. This is to be expected given that the two estimators are equivalent in this case. We shall see these trends re-occurring in all the other experiments.

When comparing the results for the two queries, we observe the confidence bounds are much tighter in the low selectivity case. Essentially, a relative error of $0.5\%$ is obtained almost from the beginning of the execution in the $8$ node configuration. The high selectivity query arrives to the same error only at the end of execution. The explanation is the $3$ orders of magnitude difference in the number of result tuples. Nonetheless, an acceptable error of $5\%$ is obtained after less than $50$ seconds ($20\%$) in the execution for the same $8$ node configuration. This is remarkable if we consider that only a fraction of $0.73\times10^{-6}$ of the overall tuples is part of the result. Although it seems more appropriate to process such a selective query using an index, the complex selection predicate makes building the correct index quite challenging. Thus, we argue that scanning the entire relation and providing estimates as in online aggregation is a more effective approach. Notice also that the execution time in all the figures is the same since the same amount of data is read from disks. In terms of estimators, there is no discernible difference between the single estimator and the multiple estimators solutions. This confirms the good accuracy of the single estimator we propose even when the data is not globally randomized.

\subsection{Group-By Aggregation}\label{ssec:exp:groupby}

\begin{figure*}
\centering
\subfloat[]{\includegraphics[width=0.5\textwidth]{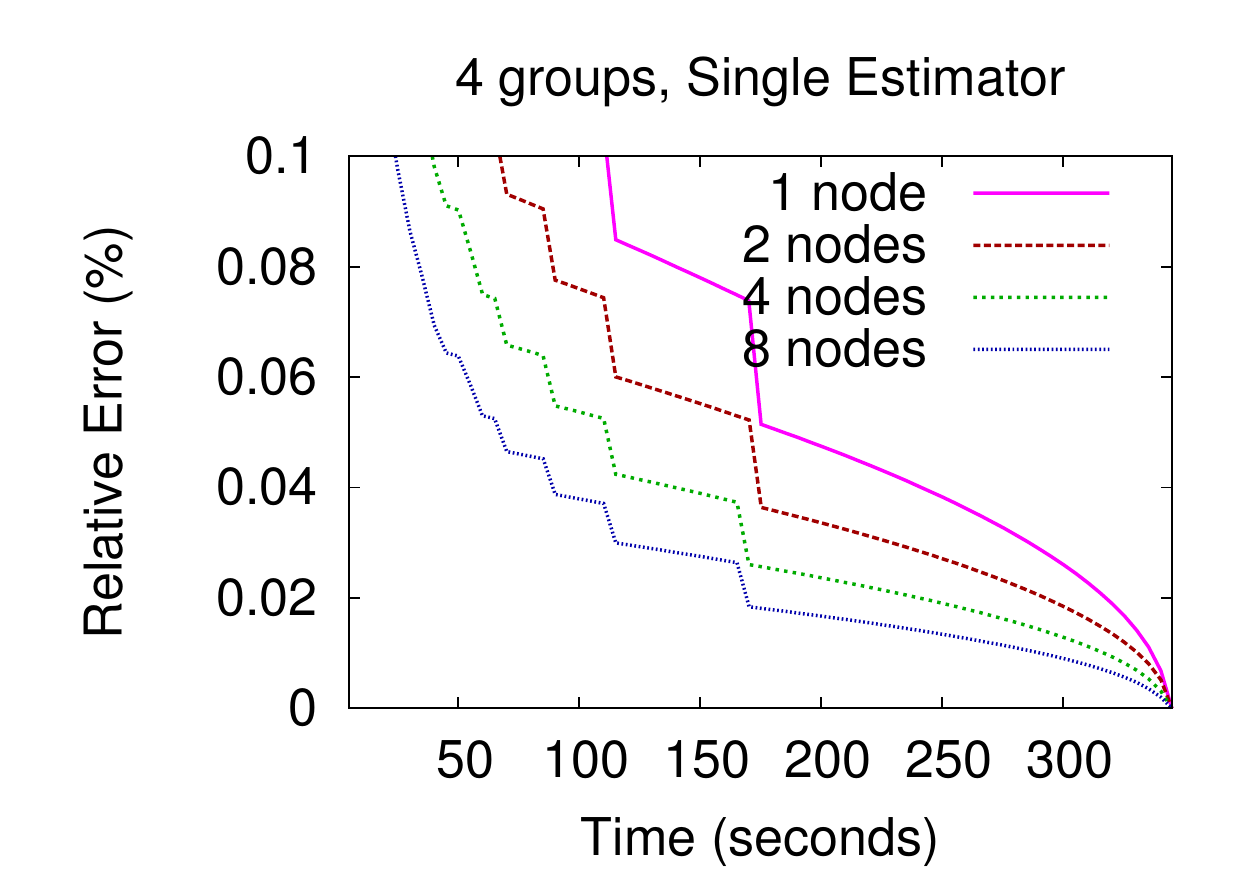}\label{fig:groupby-small-single}}
\subfloat[]{\includegraphics[width=0.5\textwidth]{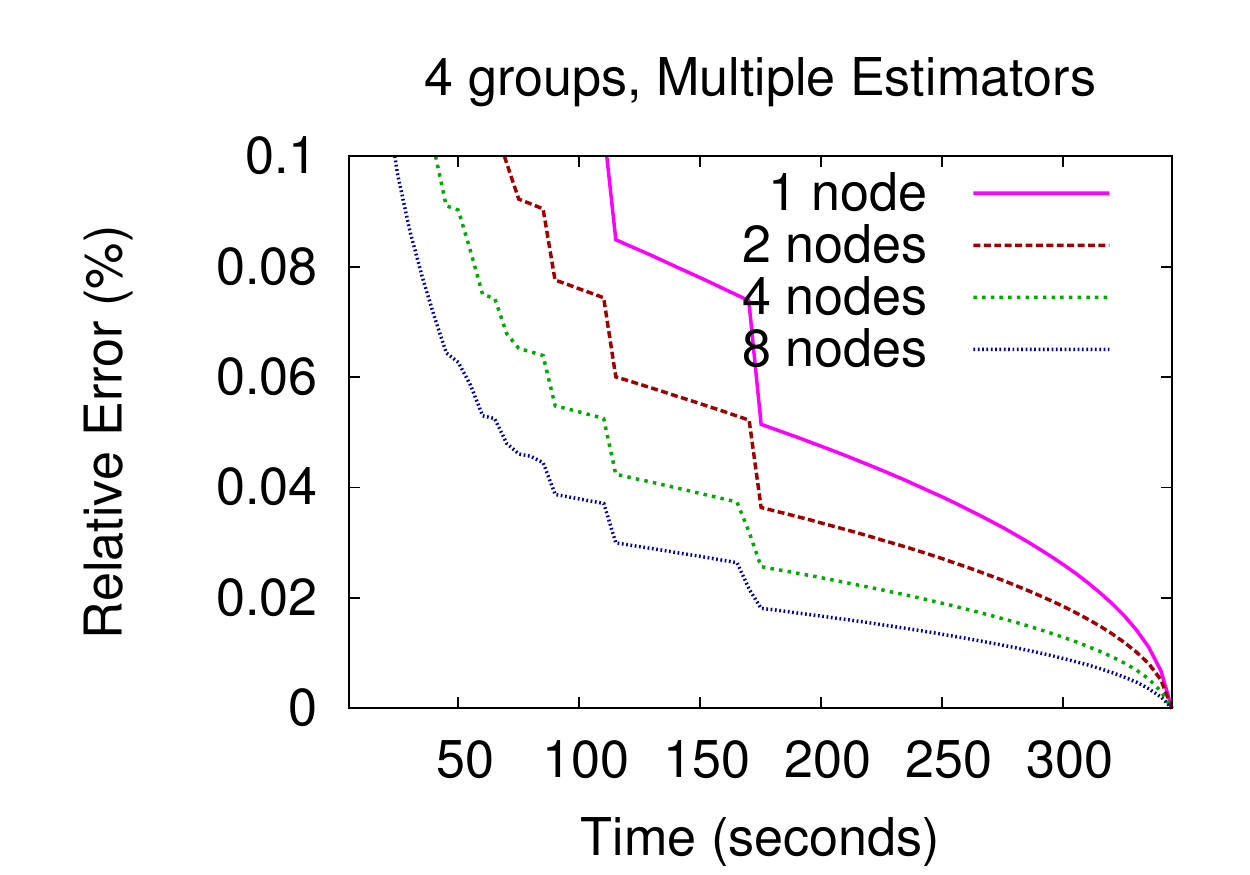}\label{fig:groupby-small-multiple}}\\
\subfloat[]{\includegraphics[width=0.5\textwidth]{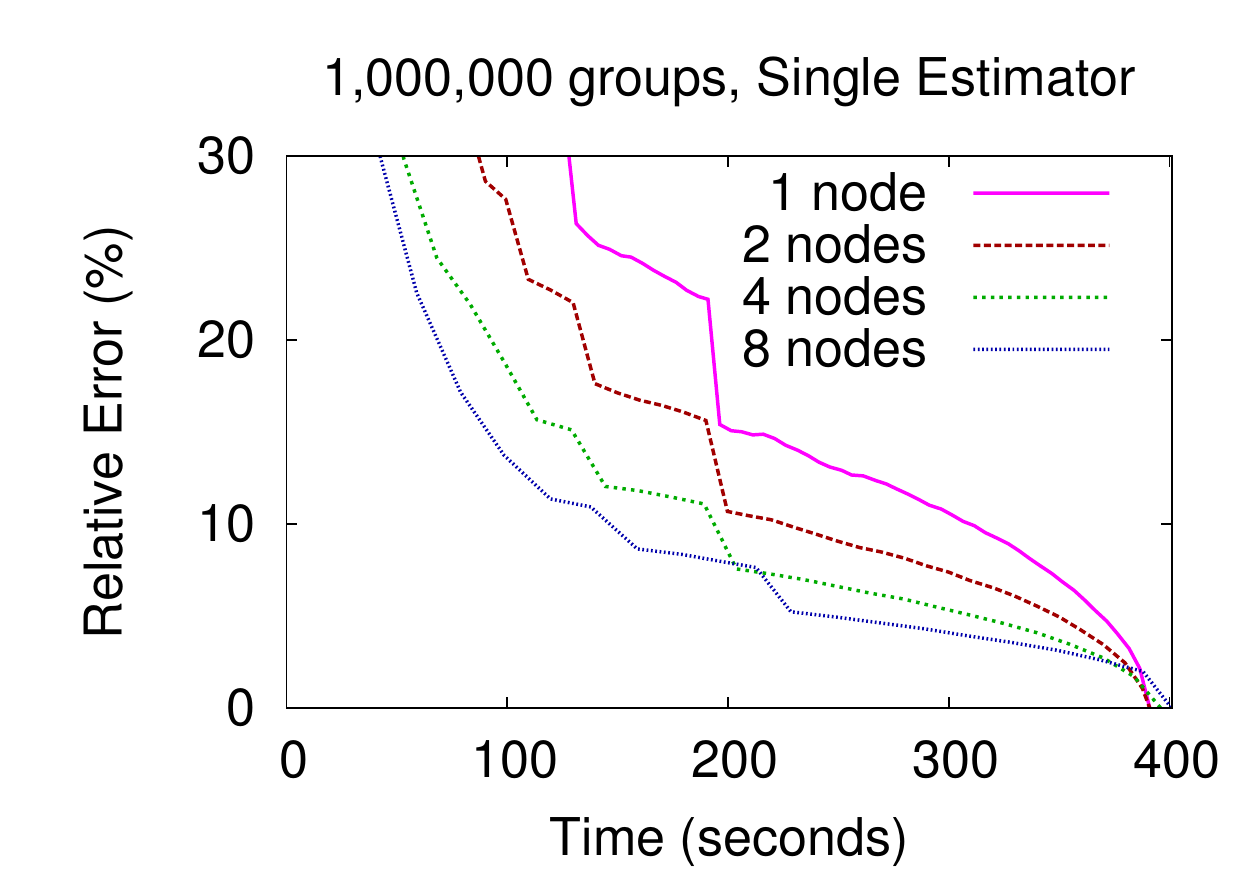}\label{fig:groupby-large-single}}
\subfloat[]{\includegraphics[width=0.5\textwidth]{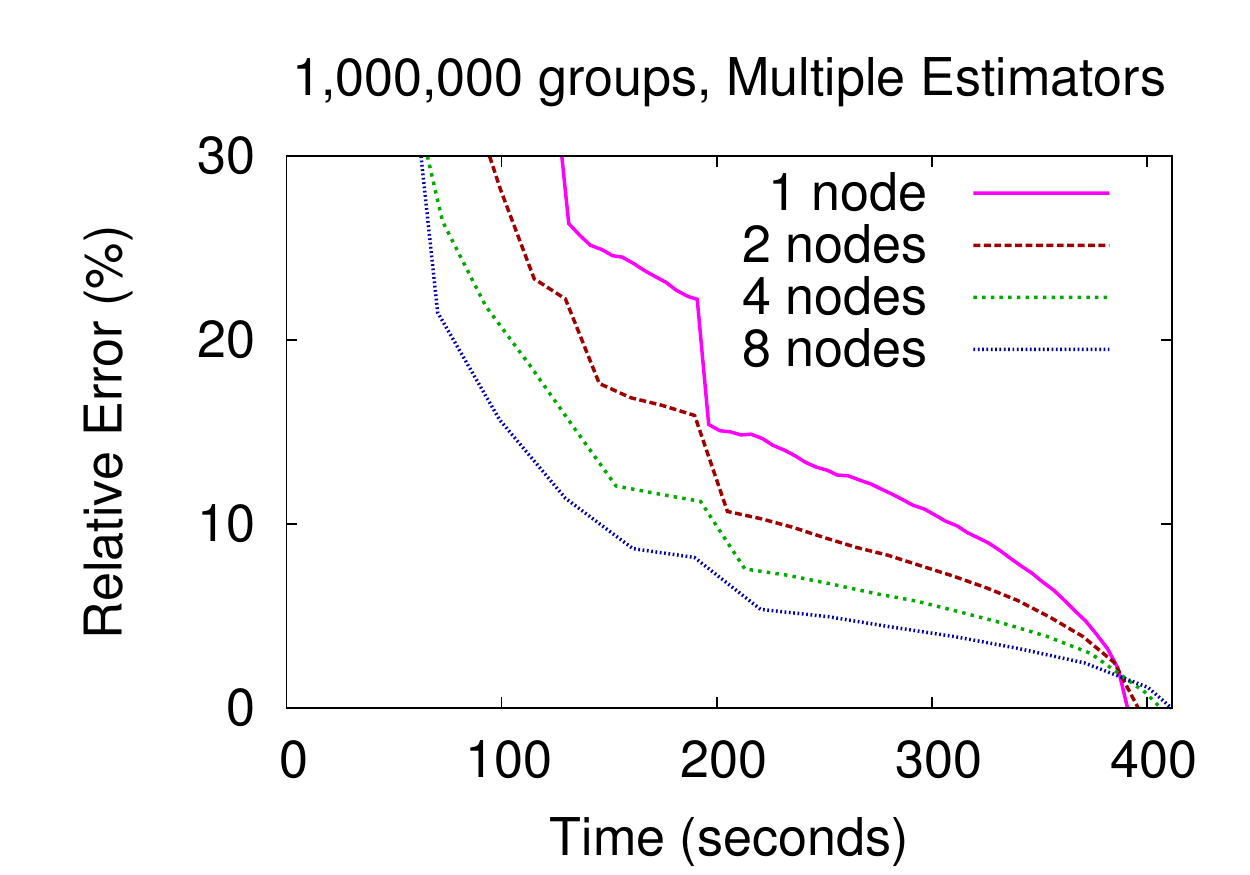}\label{fig:groupby-large-multiple}}
\caption{Comparison between single estimator and multiple estimators for the group-by task (TPC-H Q1). The figures depict relative confidence bounds width as a function of time on cluster configurations of $1$, $2$, $4$, and $8$ nodes, respectively. Figures (a) and (b) correspond to the query that generates a small number of distinct groups ($4$ to be precise), while in Figures (c) and (d) there are $1$ million distinct groups. In (a) and (b) we depict the results corresponding to the aggregate \texttt{SUM(l\_extendedprice*(1-l\_discount))} for the \texttt{`NF'} group, while in (c) and (d) the aggregate \texttt{SUM(l\_extendedprice*(1-l\_discount)*(1+l\_tax))} is shown for \texttt{l\_suppkey = `1,000,000'}, respectively. The results for the other group-aggregate combinations follow the same pattern.}
\label{fig:experiments:groupby}
\end{figure*}

We consider two group-by aggregation tasks as well. They are both instantiations of query Q1 from TPC-H:
\begin{equation*}\label{query:exp:groupby}
\setlength{\jot}{0.1pt}
	\begin{split}
		& \texttt{SELECT\hspace*{0.1cm}gAtts,\hspace*{0.1cm}SUM(l\_quantity),}\\
		& \hspace*{0.1cm}\texttt{SUM(l\_extendedprice),} \\
		& \hspace*{0.1cm}\texttt{SUM(l\_extendedprice*(1-l\_discount)),} \\
		& \hspace*{0.1cm}\texttt{SUM(l\_extendedprice*(1-l\_discount)*(1+l\_tax))} \\
		& \texttt{FROM lineitem} \\
		& \texttt{WHERE l\_shipdate between [`1998-09-01',`1998-12-01']} \\
		& \texttt{GROUP BY gAtts} \\
	\end{split}
\end{equation*}
The difference between the two queries is the grouping attributes. In the first instance we use the attributes in Q1, \texttt{l\_returnflag} and \texttt{l\_linestatus}. Since this generates only $4$ groups, we name this query \textit{group-by small}. The second instance groups by \texttt{l\_suppkey} for a total of $1$ million distinct groups, thus the name \textit{group-by large}.

Figure~\ref{fig:experiments:groupby} depicts the results for both estimators considered. Notice that estimates and bounds are computed for all group-aggregate combinations simultaneously. Only one combination is displayed. The relative confidence bounds width for group-by small follows the same pattern as for the low selectivity aggregate query. The interval width is even lower since the number of result tuples is higher. The number of tuples in the result of the group-by large query is very low ($3,464$ for the displayed group). Finding them while scanning through the entire data takes some time, thus the longer TTU. Still, a relative error of $10\%$ is obtained while only one third in the execution for the $8$ node configuration. This is remarkable given the low number of result tuples.

The increase in execution time for the group-by large query is due to the massive state of the GLA. The group-by hash table requires $1$ million distinct entries in this case, thus increasing the execution time of all operations in the UDA interface. The operation that is most affected though is \texttt{Merge}. The shorter execution time for the single node configuration certifies this. As we shall see in the runtime analysis section (Section~\ref{ssec:exp:time}) though, the overhead incurred by the estimation process is less than 1\% even in this extreme case.

\subsection{Join Group-By Aggregation}\label{ssec:exp:join}

\begin{figure*}
\centering
\subfloat[]{\includegraphics[width=0.5\textwidth]{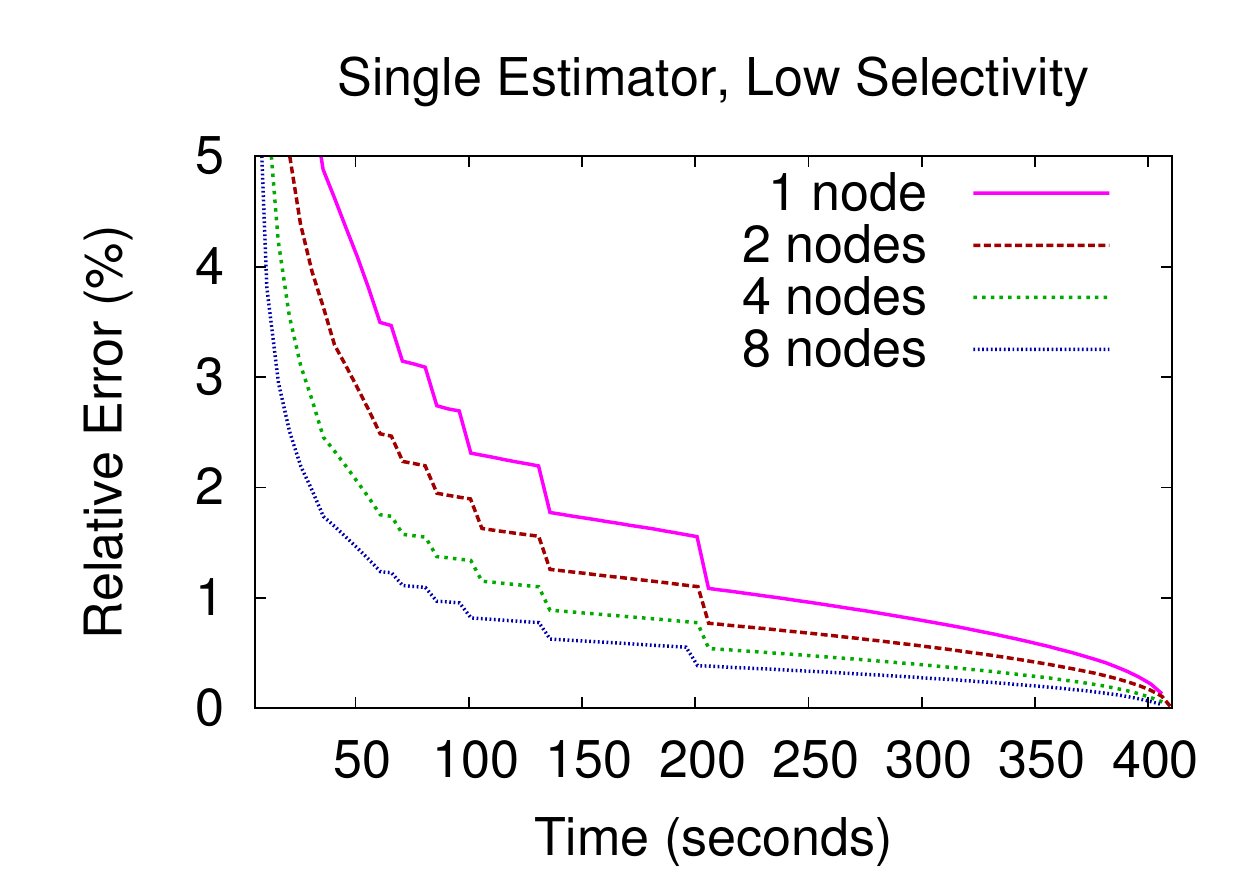}\label{fig:join-single-low}}
\subfloat[]{\includegraphics[width=0.5\textwidth]{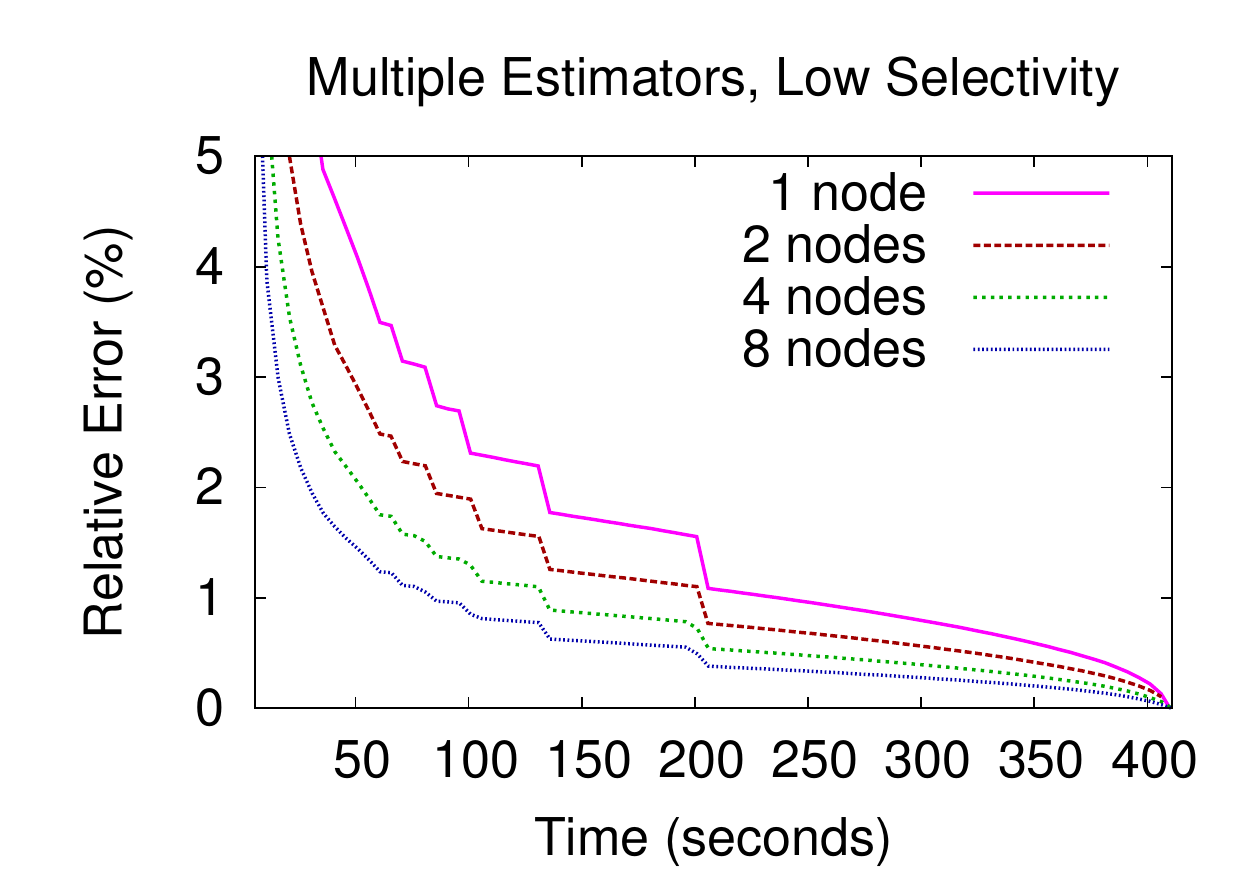}\label{fig:join-multiple-low}}\\
\subfloat[]{\includegraphics[width=0.5\textwidth]{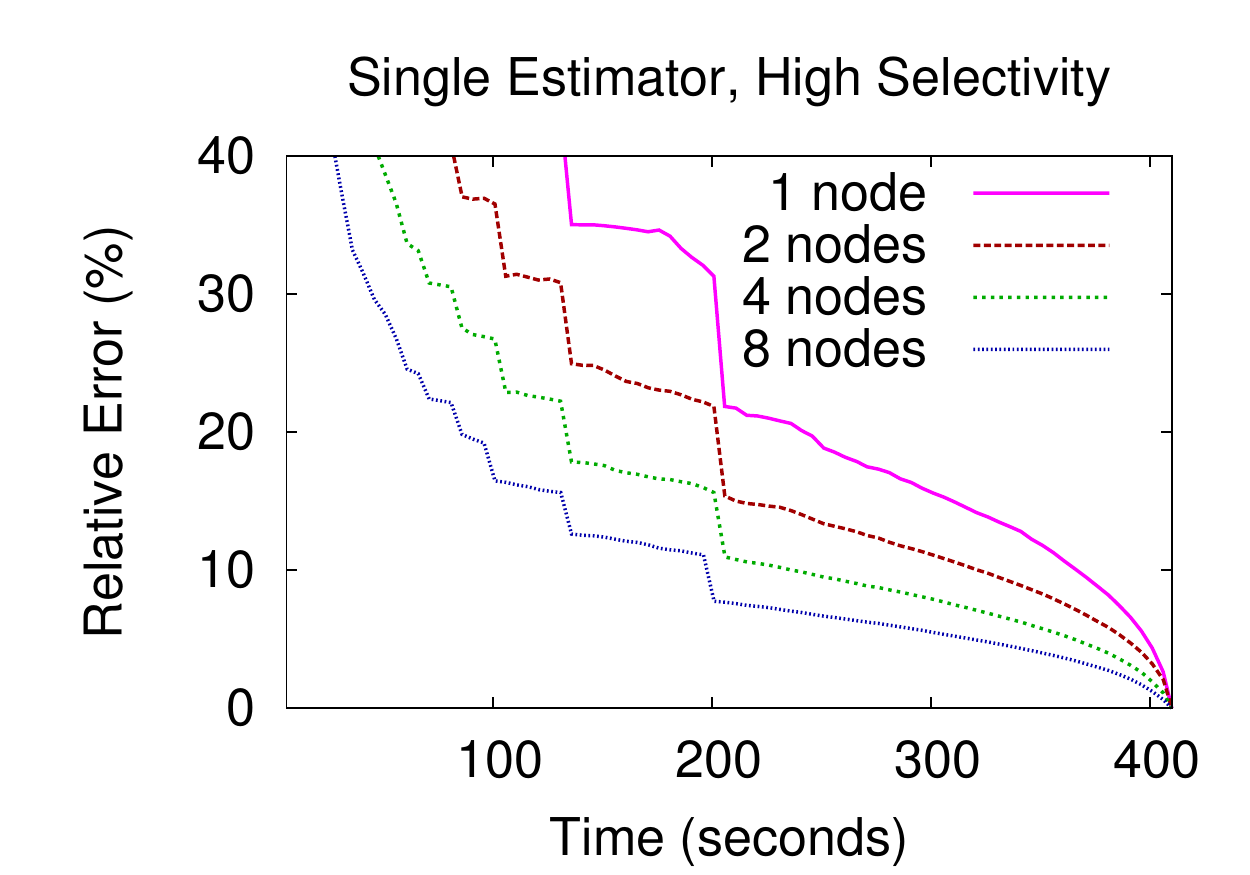}\label{fig:join-single-high}}
\subfloat[]{\includegraphics[width=0.5\textwidth]{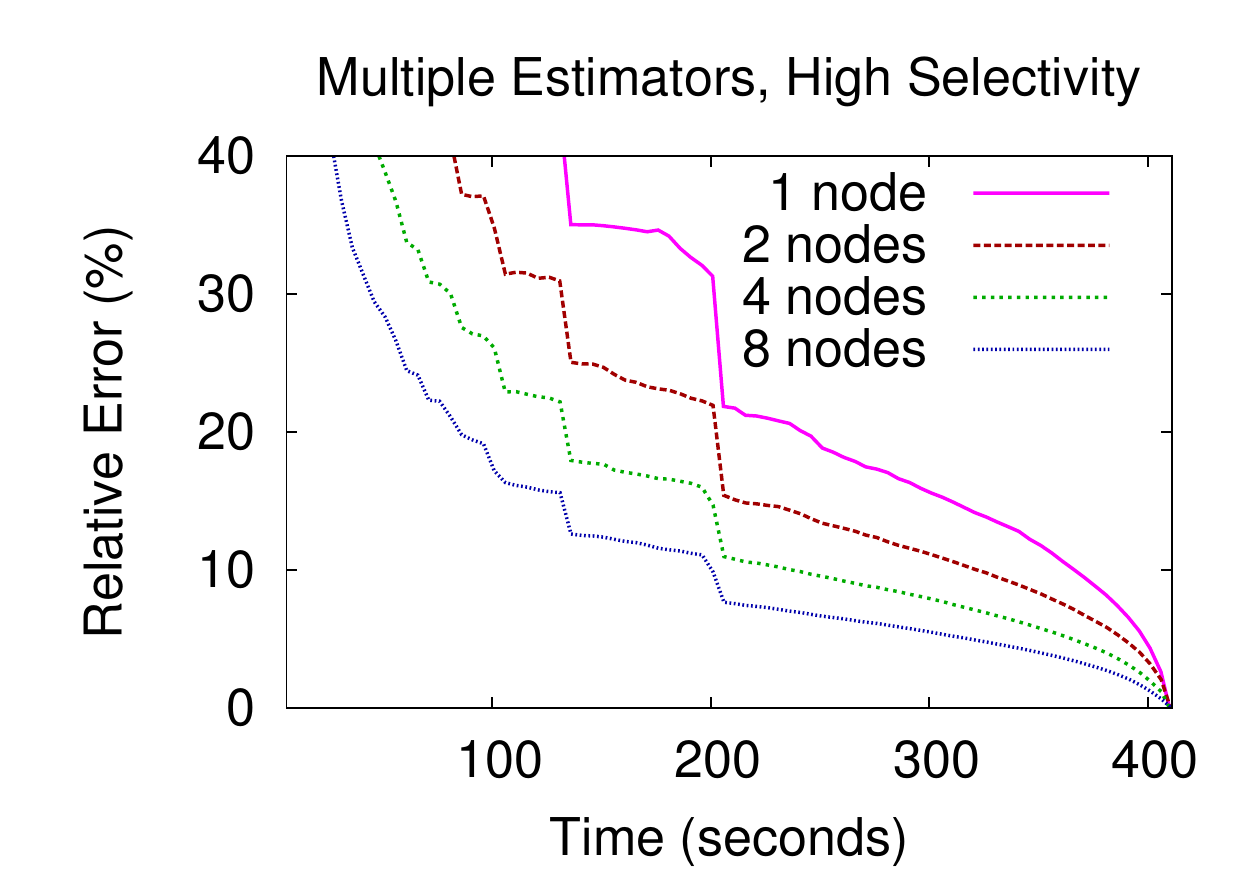}\label{fig:join-multiple-high}}
\caption{Comparison between single estimator and multiple estimators for the join group-by task. The figures depict relative confidence bounds width as a function of time on cluster configurations of $1$, $2$, $4$, and $8$ nodes, respectively. Figures (a) and (b) correspond to the low selectivity condition, while (c) and (d) correspond to high selectivity as defined in Figure~\ref{fig:experiments:agg}. The plots print the results corresponding to \texttt{SUM(l\_extendedprice*(1-l\_discount)*(1+l\_tax))} for the \texttt{`PERU'} group. The results for the other group-aggregate combinations follow the same pattern.}
\label{fig:experiments:join}
\end{figure*}

The join query we use combines the aggregate and group-by queries. It uses the selectivity conditions in the aggregate query and computes the four aggregates in the group-by query. The exact form is as follows:
\begin{equation*}\label{query:exp:join}
\setlength{\jot}{0.1pt}
	\begin{split}
		& \texttt{SELECT n\_name, SUM(l\_quantity),}\\
		& \hspace*{0.1cm}\texttt{SUM(l\_extendedprice),} \\
		& \hspace*{0.1cm}\texttt{SUM(l\_extendedprice*(1-l\_discount)),} \\
		& \hspace*{0.1cm}\texttt{SUM(l\_extendedprice\hspace*{0.00cm}*\hspace*{0.00cm}(1-l\_discount)\hspace*{0.00cm}*\hspace*{0.00cm}(1+l\_tax)\hspace*{0.00cm})} \\
		& \texttt{FROM lineitem, supplier, nation} \\
		& \texttt{WHERE l\_shipdate between [$\texttt{date}_{1}$, $\texttt{date}_{2}$] AND} \\
		& \hspace*{0.1cm}\texttt{l\_discount between [0.02, 0.03] AND} \\
		& \hspace*{0.1cm}\texttt{l\_quantity\hspace*{0.1cm}=\hspace*{0.1cm}1\hspace*{0.1cm}AND\hspace*{0.1cm}l\_suppkey\hspace*{0.1cm}=\hspace*{0.1cm}s\_suppkey\hspace*{0.1cm}AND} \\
		& \hspace*{0.1cm}\texttt{s\_nationkey = n\_nationkey} \\
		& \texttt{GROUP BY n\_name}
	\end{split}
\end{equation*}
To execute the query in parallel, \texttt{supplier} and \texttt{nation} are replicated across all the nodes. They are loaded in memory, pre-joined, and hashed on \texttt{s\_suppkey}. \texttt{lineitem} is scanned sequentially and the matching tuple is found and inserted in the group-by hash table. Merging the GLA states proceeds then as in the group-by case. This join strategy is common in parallel databases.

The results are depicted in Figure~\ref{fig:experiments:join}. A similar trend as in Figure~\ref{fig:experiments:agg} can be observed. This is expected given the similar selectivity conditions. The minor differences are due to the number of tuples in the result. What is remarkable though is the reduced TTU even for extremely selective queries with only $1,000$ tuples in the result (Figure~\ref{fig:join-single-high} for example). This is a very extreme case where there are no result tuples in many chunks and we still achieve lower than $10\%$ relative error before half of the computation is executed. As with the other queries, there is no discernible difference between single estimator and multiple estimators.

\subsection{Data Skew}\label{ssec:exp:zipf}

\begin{figure*}
\centering
\subfloat[]{\includegraphics[width=0.5\textwidth]{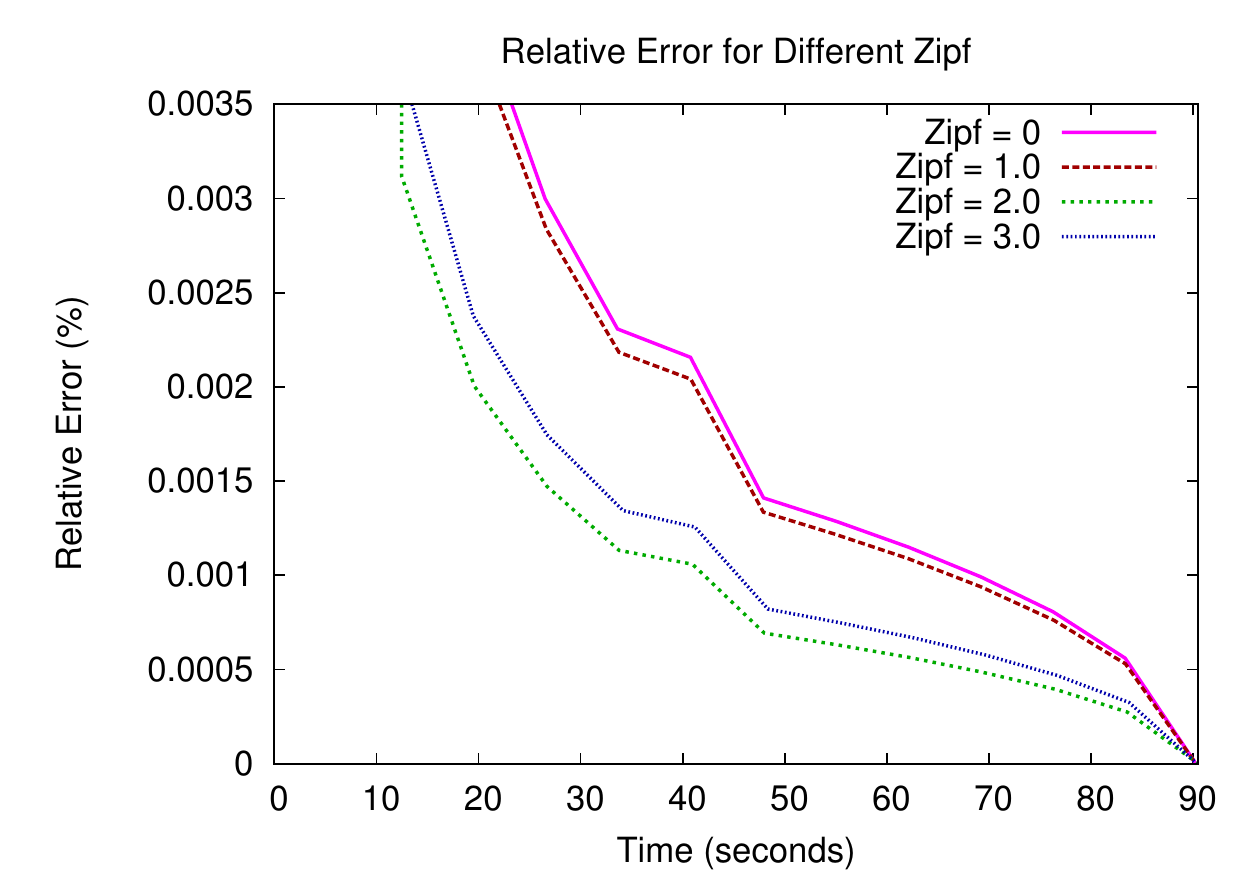}\label{fig:zipf-single}}
\subfloat[]{\includegraphics[width=0.5\textwidth]{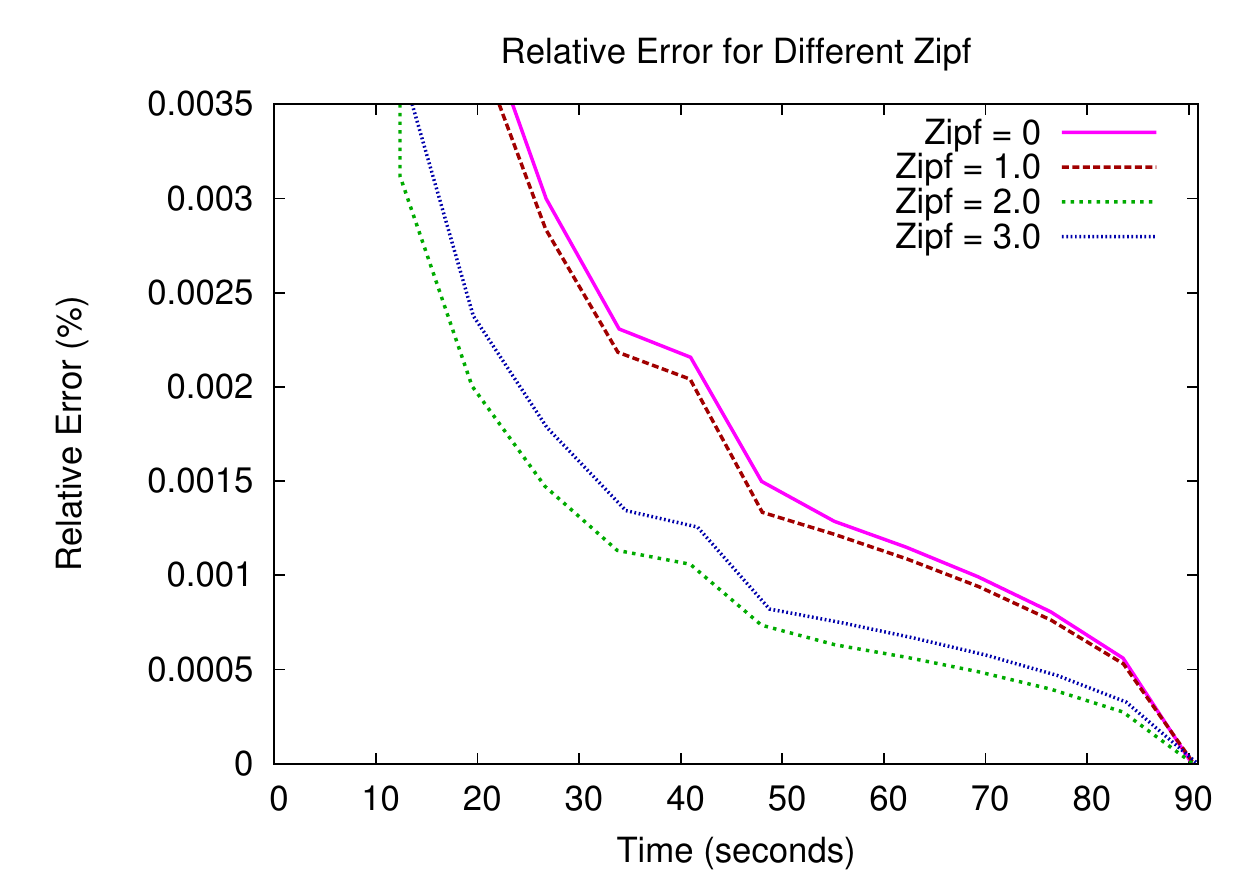}\label{fig:zipf-multiple}}\\
\subfloat[]{\includegraphics[width=0.5\textwidth]{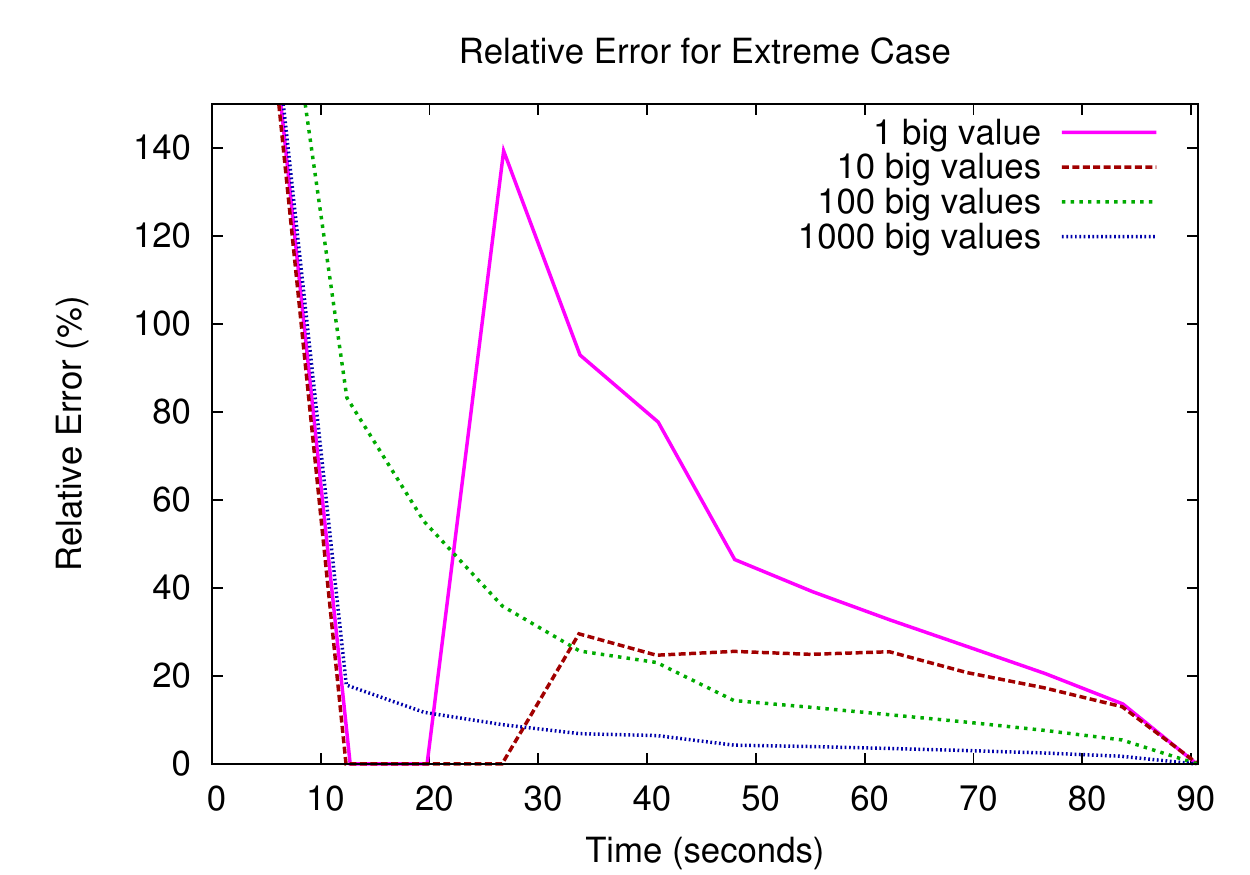}\label{fig:extreme-single}}
\subfloat[]{\includegraphics[width=0.5\textwidth]{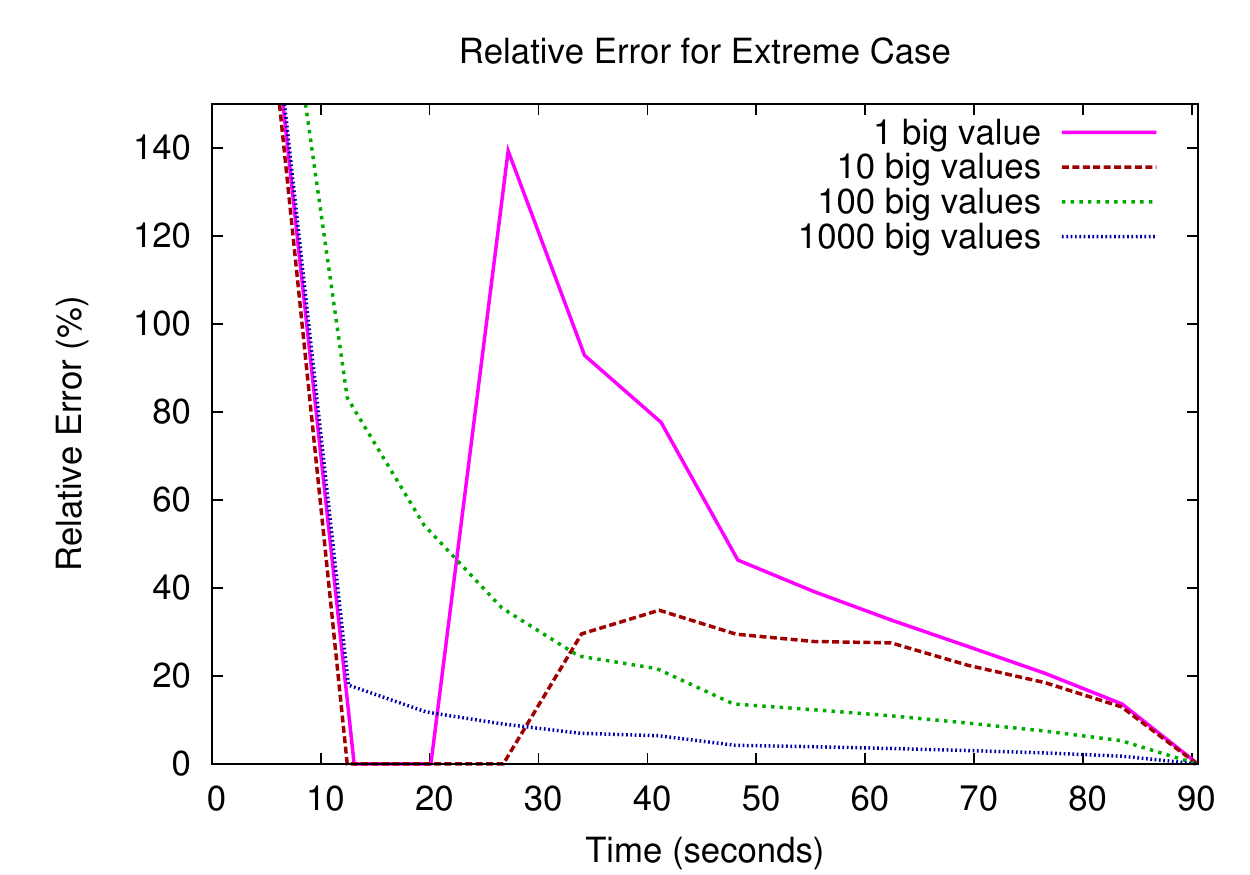}\label{fig:extreme-multiple}}
\caption{Effect of data skew on the accuracy of the estimators. (a) and (b) depict the accuracy for datasets generated with different Zipf distributions. (c) and (d) correspond to extreme-case datasets where a few values are orders of magnitude larger than all the other values.}
\label{fig:experiments:skew}
\end{figure*}

It is a well-known fact that the TPC-H dataset exhibits a distribution close to uniform. Someone might argue this is the reason why the accuracy of the estimators in the previous experiments is so high. While this argument holds for the low selectivity queries, it is not true for the queries with extremely low selectivity.

To further verify this, we run a series of experiments in a tightly controlled setting meant to study the accuracy as a function of the data skew. For this, we generate a dataset consisting of $3.2\times 10^{9}$ items distributed over a domain with $10^{6}$ elements using a zipfian distribution with different parameters. The assignment of frequency to domain elements is random. Global randomization is applied to distribute data across the 8 nodes.

Figure~\ref{fig:zipf-single} and \ref{fig:zipf-multiple} depict the estimator accuracy as a function of the zipfian parameter for the aggregation task. Although we might expect some kind of sensitivity to the skew in the data, the estimators seem to not be affected by the skew. At first sight, this seems unreasonable. After a careful analysis of the experimental conditions, we found the explanation. The first estimator is already computed over $6\%$ of the data. This represents approximately $200\times 10^{6}$ tuples---a massive sample. At such a sample size, no matter how large and skewed are the data, the accuracy is going to be high. What makes PF-OLA special is that extracting such a large sample takes only a few seconds. It is practically impossible to get small samples due to the push-based processing implemented at the lowest level of the system.

In the experiments depicted in Figure~\ref{fig:extreme-single} and \ref{fig:extreme-single}, we test the accuracy of the estimators for pathological data distributions. Out of the $3.2\times 10^{9}$ tuples, all take value $1$, except for a handful of outliers which take value $10^{9}$. Sampling accuracy in such situations is known to be extremely poor. The figures confirm this to be the case but only for the most extreme situations where only $10$ and, respectively, $1$ values are outliers. We do a simple computation to confirm these results. For $100$ outliers, the first sample already contains $6$ large values with high probability---the exact positions depend on the global randomization. This is more than enough for high accuracy.

\subsection{Robustness}\label{ssec:exp:robust}

\begin{figure*}
\centering
\subfloat[]{\includegraphics[width=0.5\textwidth]{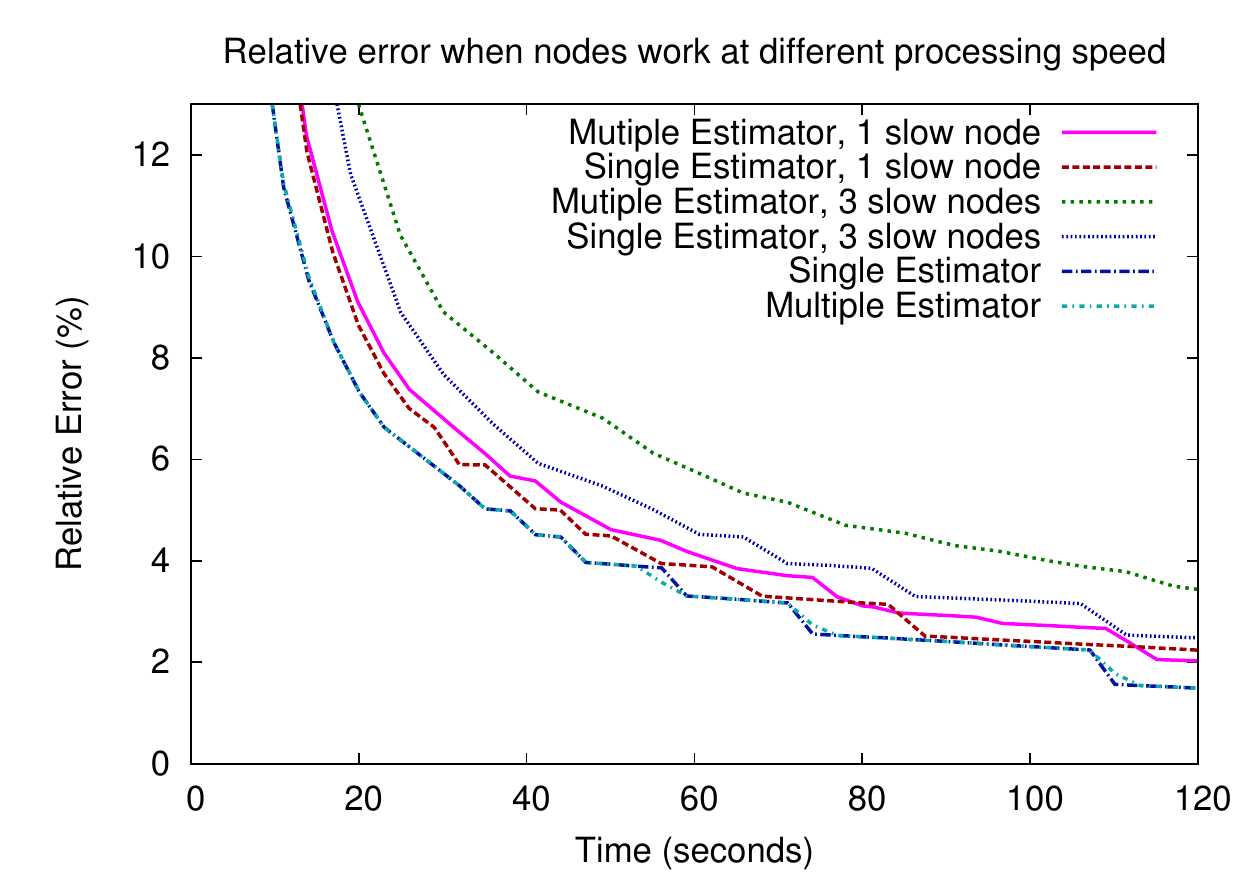}\label{fig:straggler}}
\subfloat[]{\includegraphics[width=0.5\textwidth]{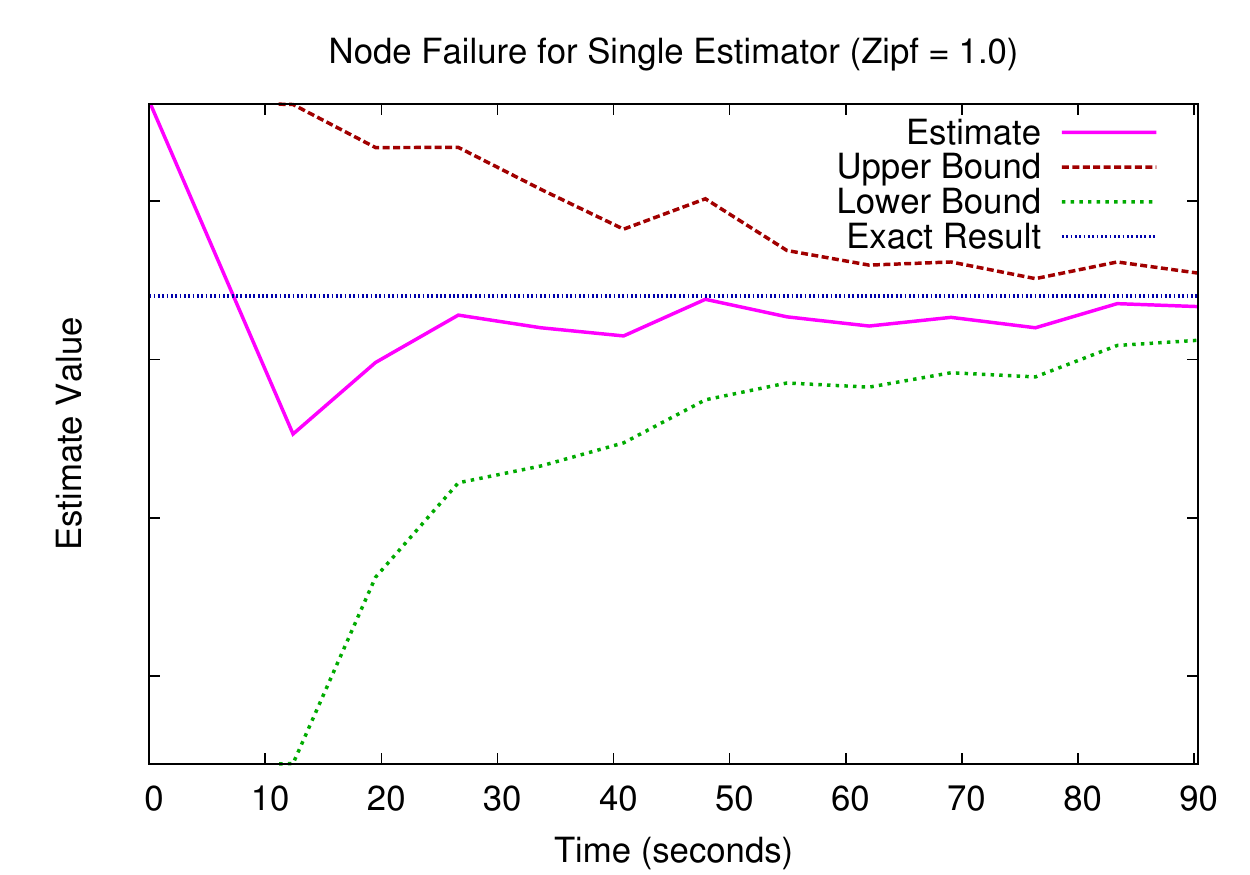}\label{fig:dead}}
\caption{(a) Effect of straggler node(s) on estimation. (b) Effect of "dead" node(s) on the single estimator with global randomization.}
\label{fig:exp:robust}
\end{figure*}

We validate the robustness of the estimators presented in the paper in two different scenarios. In the first scenario, one or multiple nodes in the cluster process data at a considerable slower rate than the other nodes---they are stragglers. In the second scenario, one node "dies" and the corresponding data are not available anymore---data are not replicated across nodes.

Figure~\ref{fig:straggler} depicts the relative errors in the scenario where one or three nodes (out of 8) work at slower speed than others and compares the results with the normal configuration when all the nodes work at roughly the same speed. While relative errors are almost the same for the two estimators in the normal case, the curves start to separate when we introduce delays at some of the nodes. We delay a node by intentionally pausing several seconds for every chunk processed. The delays are different across nodes. As a result, the sample size of the delayed node(s) is considerably smaller. This results in the decrease of the accuracy for both estimators. However, single estimator has better accuracy than multiple estimators even though the overall sample size is the same. What is different though is the distribution of the samples across nodes. In the single estimator case, the overall sample increases at a lower rate, thus the accuracy is worse than in the normal configuration. For multiple estimators, different sample sizes at nodes affect only the local estimator. While this might not be considerable for a single node taken separately, the effect is amplified when all the nodes are considered together. Concretely, the variance of the slow nodes is larger. When adding all the variances together to obtain the overall variance for the multiple estimators, we end up with a larger value than the variance of the single estimator computed from a smaller sample size.

In the second scenario, we consider only seven nodes in the estimation process. We still want to estimate the result over all the data, including the missing partition. For multiple estimators, this is simply not possible because we cannot get any estimate on the missing data. In the single estimator case, samples are always drawn from the overall data, independent of the number of available partitions. In this particular case, the maximum sample size is limited to the available data. As a result, the confidence bounds do not collapse on the true result anymore even at the end of query execution (Figure~\ref{fig:dead}).

\subsection{Correctness}\label{ssec:exp:monte-carlo}

\begin{figure*}
\centering
\subfloat[]{\includegraphics[width=0.5\textwidth]{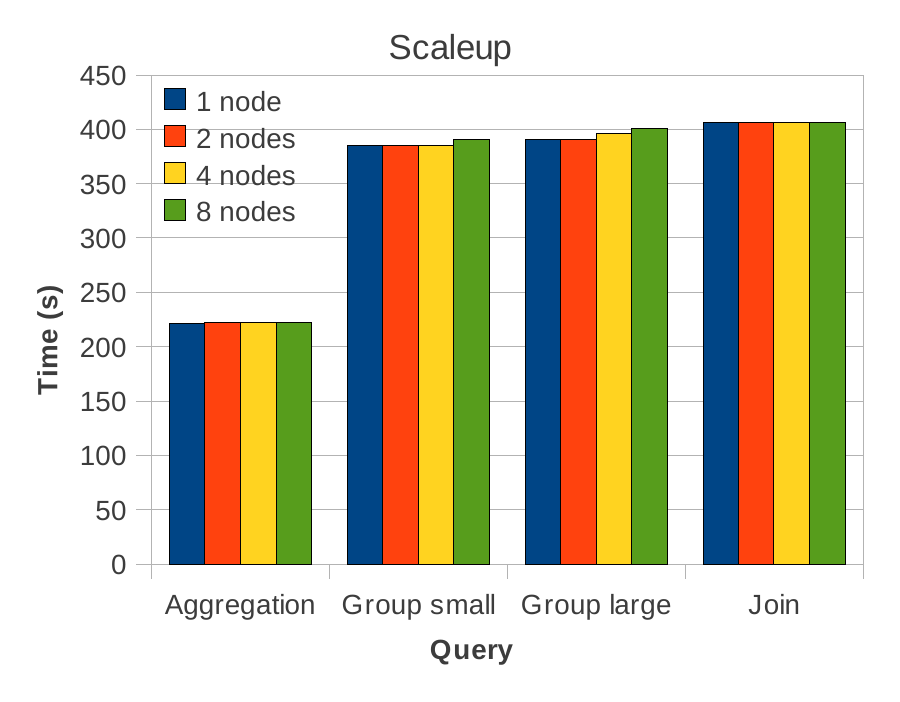}\label{fig:scaleup}}
\subfloat[]{\includegraphics[width=0.5\textwidth]{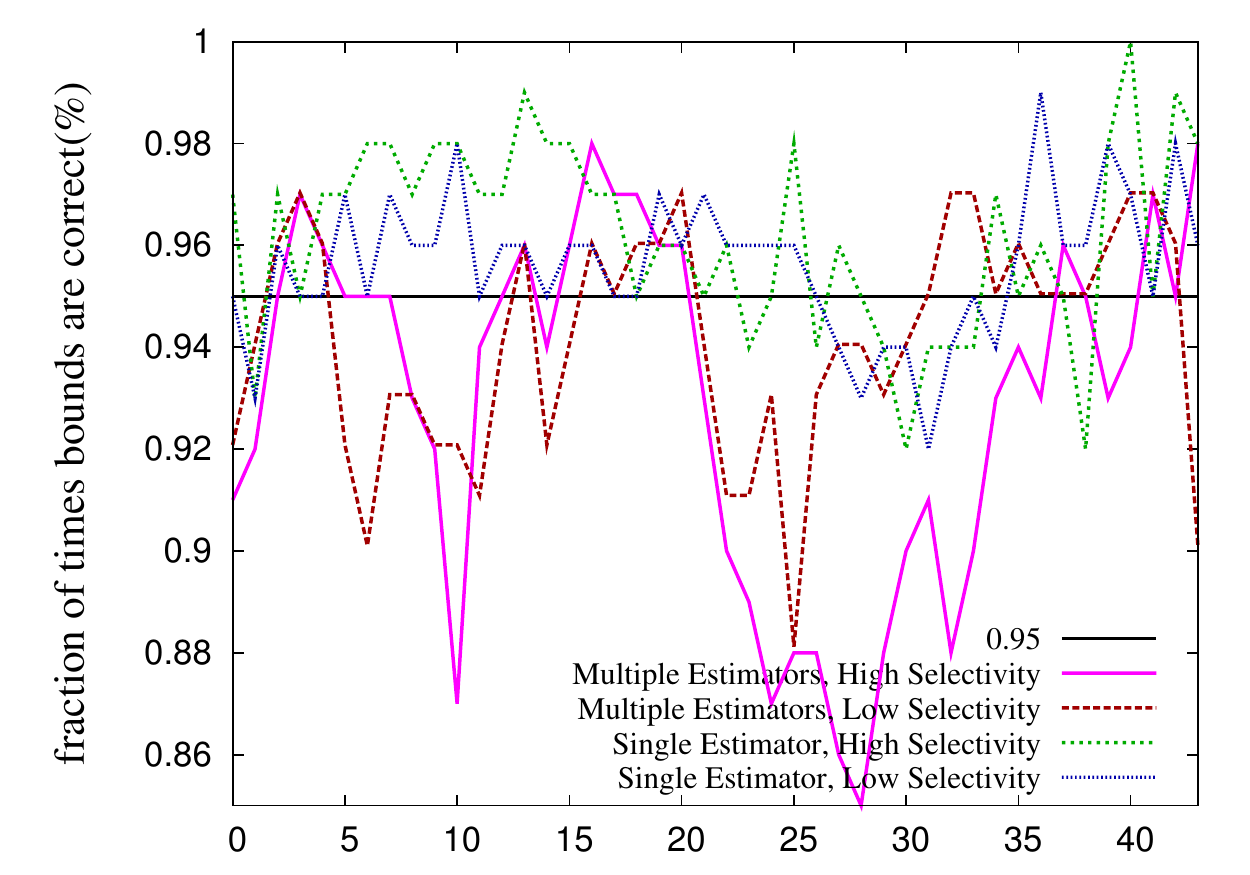}\label{fig:exp:MC}}
\caption{(a) Query execution scaleup. (b) Monte Carlo simulations to check confidence bounds correctness.}
\label{fig:exp:MC+scale}
\end{figure*}

To verify the correctness of the confidence bounds produced by the two estimators, we execute $100$ Monte Carlo trials with different data randomizations for the aggregation queries. The $8$ node configuration is used. Figure~\ref{fig:exp:MC} plots the number of times the correct result is between the estimated confidence bounds at $5$ second intervals during query execution (there are $43$ such points over $222$ seconds). The confidence level is set at $95\%$, thus the additional horizontal line. From the figure, we observe that only the bounds for the single estimator approach verify the required confidence level (they are centered around the $0.95$ horizontal line). Multiple estimators has a considerable drop during the second part of the query execution. The reason for this is the asynchrony in execution that starts to affect the estimator. Though we use the same configuration across all the nodes, the minor difference in the number of samples that have been processed at each node starts to accumulate and finally leads to an increasing difference in estimates among nodes. Thus, the multiple estimators method is more prone to generate errors when we combine the estimates together according to the stratified sampling theory~\cite{sampling-techniques} since stratified sampling is not in the optimal regime---the samples at nodes are not proportional to the size of the strata. This behavior also proves that the single estimator we propose is more stable and a better solution for parallel online aggregation. We further confirm this with experiments that test the robustness of the estimators.

\subsection{Speed \& Scalability}\label{ssec:exp:time}

Table~\ref{tbl:exec-time} contains the execution time for the $8$ node configuration. It confirms that online aggregation can be enabled for any query at virtually no overhead. As far as we know, PF-OLA is the first online aggregation system achieving this goal. It is the asynchronous overlapping of execution and estimation in the GLADE parallel multi-query processing system that makes this possible. If the synchronized single estimator proposed in~\cite{distributed-ola} is used, the execution time for the aggregation task is $883$ seconds, a factor of $4$ increase on top of the normal execution. Clearly, this is not acceptable since our goal is to not increase the execution time beyond normal execution. For this reason, we do not even include the results for the other tasks in Table~\ref{tbl:exec-time}.

\begin{table}[htbp]
    \begin{center}
      \begin{tabular}{|p{3.cm}|p{3.cm}|p{2.7cm}|p{3.cm}|}

    \hline
	\multirow{2}{*}{\textbf{Query}}  & \multicolumn{3}{|c|}{\textbf{Execution Time (seconds)}}  \\ \cline{2-4}

	 & \texttt{No estimation} & \texttt{Single estimator} & \texttt{Multiple estimators} \\
	\hline\hline	
	\texttt{Aggregate} & \multicolumn{1}{|r|}{222} & \multicolumn{1}{|r|}{222} & \multicolumn{1}{|r|}{222} \\
	\hline\hline
	$\texttt{Group}_{\texttt{small}}$ & \multicolumn{1}{|r|}{344} & \multicolumn{1}{|r|}{345} & \multicolumn{1}{|r|}{344} \\
	\hline	
	$\texttt{Group}_{\texttt{large}}$ & \multicolumn{1}{|r|}{404} & \multicolumn{1}{|r|}{407} & \multicolumn{1}{|r|}{407} \\
	\hline\hline
	\texttt{Join} & \multicolumn{1}{|r|}{409} & \multicolumn{1}{|r|}{411} & \multicolumn{1}{|r|}{411} \\
	
	\hline

      \end{tabular}
    \end{center}

    \caption{Execution time.}
    \label{tbl:exec-time}
\end{table}

To clarify the surprisingly low execution time, we analyze in detail the time it takes to execute the group-by small query. Remember that we have TPC-H scale $1,000$ data on each node. \texttt{lineitem} has $6\times10^{9}$ tuples for this instance. This corresponds to $1.5\times10^{9}$ tuples per disk. DataPath uses columnar storage, thus it reads only the columns required by the query. In this case it reads $7$ columns, summing-up to $28$ bytes per tuple and $42\textit{GB}$ per disk. Given the $130\textit{MB/s}$ disk bandwidth, the theoretical execution time is $323$ seconds. This comes close to the actual execution time of $344$ seconds, but is not exactly the same. Here are the reasons. We monitored the disk bandwidth during the execution and it was only $123\textit{MB/s}$. According to \texttt{iostat} the disk was fully utilized, thus the execution was I/O-bound. With this bandwidth, the execution time is $341.5$ seconds. We already knew that the difference is taken by the time to setup the query, thus everything makes perfect sense. It is possible to run similar analyses for the other queries. More parameters such as the time to merge two GLAs, the time to serialize/deserialize the GLA state, and the time to transfer the data between the nodes might need to be accounted for though.

Although the scaleup of PF-OLA can be inferred from the accuracy figures -- the execution time corresponding to all the configurations is the same -- Figure~\ref{fig:scaleup} shows explicitly the time it takes to run each task on 1, 2, 4, and 8 nodes when the size of the data increases with a corresponding factor. Since the execution time remains constant, this confirms that PF-OLA scales-up linearly.

\subsection{Discussion}\label{ssec:exp:discuss}

The main findings of the experimental study are:
\begin{compactitem}
\item The study proves the expressiveness of the PF-OLA framework. Two estimation models are created for three different tasks, each consisting of two queries. They are all implemented as GLAs with the extended UDA interface and executed successfully by the framework. 
\item The proposed single estimator has similar TTU and identical or better accuracy than multiple estimators. PF-OLA is able to provide accurate estimations for the query result early in the execution even in the most difficult scenarios when only a handful of tuples are part of the result or when data are skewed.
\item The asynchronous single estimator we propose is highly insensitive to processing node delays and failure. This is the signature characteristic when compared to the multiple estimators solution.
\item No significant overhead is incurred by online aggregation for any of the queries. The execution is always I/O-bound.
\item PF-OLA has perfectly linear scaleup as the data and the processing resources increase proportionally.
\item The correctness of the estimators is confirmed through Monte Carlo simulations.
\end{compactitem}

\section{Related Work}\label{sec:rel-work}

There is a plethora of work on online aggregation published in the database literature~\cite{AQP-book} starting with the seminal paper by Hellerstein et al.~\cite{ola}. We can broadly categorize this body of work into system design~\cite{control,dbo,demo:dbo,turbo:dbo}, online join algorithms~\cite{ripple-join,sms-join,pr-join}, online algorithms for estimations other than join~\cite{ola-set,fei-groupby,ola-extreme}, and methods to derive confidence bounds~\cite{haas-bounds}. All of this work is targeted at single-node centralized environments. The parallel online aggregation literature is not as rich though. We identified only a relatively small number of research papers that are closely related to our work. We discuss them in details in the following.

Luo et al.~\cite{scalable-hash-ripple} extend the centralized ripple join algorithms~\cite{ripple-join} to a parallel setting. The proposed parallel hash ripple join algorithm is a non-blocking version of the parallel hybrid hash join algorithm that allows for estimates to the final query result to be computed. A stratified sampling estimator~\cite{sampling-techniques} is defined to compute the result estimate while confidence bounds cannot always be derived. We implement a similar stratified sampling estimator in PF-OLA and compare it with the estimator we propose. Our focus is on analyzing the properties of the two estimators along a larger set of dimensions, including robustness, which is not discussed at all in~\cite{scalable-hash-ripple}. Moreover, the prototype system implementing the specific parallel hash ripple join algorithm is very particular to the proposed estimator. There is no common framework proposed for general parallel online aggregation.

Wu et al.~\cite{distributed-ola} extend online aggregation to distributed point-to-point (P2P) networks. They introduce a synchronized sampling estimator over partitioned data that requires data movement from storage nodes to processing nodes. We also implement this estimator in PF-OLA and show the poor performance it achieves in a highly-parallel asynchronous system. In subsequent work~\cite{continuous-sampling}, Wu et al. tackle online aggregation over multiple queries. They emphasize the benefits of global randomization as a sample generating method.

Recently, online aggregation in Map-Reduce emerged as a popular research area. This is mostly motivated by the poor performance of the Map-Reduce implementation in Hadoop~\cite{hadoop}, which makes online aggregation a necessity rather than a luxury in the context of Big Data. Hadoop Online (HOP)~\cite{hadoop-online} is an extension to the Hadoop framework which allows for partial aggregates to be extracted during execution using pipelining between operators. Stock Hadoop contains only blocking operators that materialize the intermediate results for fault-tolerance. As explained in Section~\ref{ssec:par-online-partial-agg}, partial result extraction is only the basic requirement for online aggregation. Sampling and estimation provide significance to the partial results. HOP is limited only to partial result extraction. There is no formal sampling or estimation involved. Thus, HOP is not an online aggregation system. It is a partial aggregation system.

It is the work in~\cite{online-mapreduce} where HOP is elevated to a real online aggregation system by providing an estimation mechanism. The proposed solution is a Bayesian framework to handle the correlation between the time to process a data partition and the result it generates. This is required because chunks are treated as black boxes that only produce an aggregate. There is no information on what operation was performed to generate the aggregate or on the content of the chunk. Based on the aggregates produced by the processed chunks and the time it took to schedule and process the chunk, a prediction is made for the aggregates in the chunks not scheduled yet---the processed chunks are an independent and identically distributed (iid) sample from the entire chunk population. The Bayesian model is continuously updated as more chunks are processed. This results in more accurate estimates as more data are processed. Although this estimation model is not based on sampling, it can still be expressed as a GLA using the extended UDA interface. We plan to do this in future work to further validate the expressiveness of the PF-OLA framework. 

BlinkDB~\cite{blinkdb} stores pre-computed samples of different sizes on disk. This requires a significant amount of additional storage on top of the original data which might be a problem if the dataset is massive. Moreover, the time to compute the samples -- even if executed offline -- might be prohibitive. Similar to iterative online aggregation, a query is evaluated on a chosen sample and an estimate is produced. If the accuracy is not satisfying, a subsequent query can be executed on a larger sample---in the worst case, the query is executed on the entire dataset. While this allows for discrete estimates of increasing accuracy, it cannot support continuous estimation. Determining the correct sample to execute the query on is a complicated problem that requires estimating the variance of the result. None of these problems arise in PF-OLA as long as global randomization is executed on the data. This is a considerably less time-consuming process than taking samples of progressively increasing sizes. A somehow similar idea is used in EARL~\cite{earl} where a single sample is pre-computed. Bootstrapping is then used to extract multiple samples from the pre-computed sample and compute estimates. The sample sizes are increased dynamically to provide better accuracy. Different from BlinkDB, if the large sample in EARL has to be re-computed, this is done dynamically at runtime.

Different estimation algorithms are proposed in each reference we mention. No common framework for estimation exists. PF-OLA provides a common framework to model a much larger class of estimation models. In terms of performance, all the estimation methods incur considerable overhead. The only exception is PR-Join~\cite{pr-join} which combines a non-blocking join algorithm with temporary storage on solid-state drives (SSD) to produce the result tuples much faster, thus increasing the convergence rate. It is a centralized algorithm though.

\section{Conclusions}\label{sec:conclusions}

We present PF-OLA, the first framework for parallel online aggregation that does not incur any noticeable overhead on top of the actual computation. The extensive use of parallelism at all levels of the system and the sound overlapping between computation and estimation make this possible. PF-OLA provides an abstract interface enhancing the well-known UDA to express any estimation model. The framework handles all the execution details in a parallel environment allowing the analyst to focus on estimation. We design a novel asynchronous sampling estimator for parallel online aggregation over partitioned data. While achieving similar accuracy to existing estimators, our estimator is much more robust to node delays and failures. To verify the capabilities of the framework, we compare our estimator with two existing estimators by implementing and executing them in the framework. The results confirm the ability of the framework to execute the estimators without incurring any remarkable overhead and to provide tight confidence bounds early in the execution for highly selective queries, skewed data, and other pathological cases. The reason for this is the extremely efficient tuple discovery mechanism that takes advantage of multi-node and multi-threaded parallelism.

We plan to address some of the limitations of the framework and extend its capabilities in future work. We have already started to investigate how to provide online estimates for asynchronous parallel joins when none of the two relations fits in memory. We plan to incorporate other estimation methods than sampling in the framework, for example Bayesian statistics and bootstrapping. Our long-term goal is to provide online estimates for any computation without incurring any overhead. We believe this is possible given the amount of parallelism available in modern processors.

\bibliographystyle{abbrv}

\end{document}